\documentclass[10pt]{article}
\usepackage{sao2}
\usepackage{psfig,epsf}

\issue{2010, 65, 238--249}

\def\apj{Astrophys. J}

\def\apjs{Astrophys. J. Supp.}

\begin{document}
\markboth{O.V.~Verkhodanov,V.V.~Sokolov, M.L.~Khabibullina, S.V.~Karpov}
  {GRB SKY DISTRIBUTION PUZZLES}
\title{GRB Sky Distribution Puzzles}
\author{O.V.~Verkhodanov\inst{a}
V.V.~Sokolov\inst{a}
M.L.~Khabibullina\inst{a}
S.V.~Karpov\inst{a}
}
\institute{
\saoname}

\date{February 25, 2010}{April 21, 2010}
\maketitle

\begin{abstract}
We analyze the randomness of the sky distribution of cosmic
gamma-ray bursts. These events are associated with massive
galaxies, spiral or elliptical, and therefore their positions
should trace the large-scale structure, which, in turn, could show
up in the sky distribution of fluctuations of the cosmic microwave
background (CMB). We test this hypothesis by mosaic correlation
mapping of the distributions of CMB peaks and burst positions,
find the distribution of these two signals to be correlated, and
interpret this correlation as a possible systematic effect.

\end{abstract}

\maketitle

\section{INTRODUCTION}

The publication of comparatively large catalogs of gamma-ray
bursts based on the results of the surveys performed within the
framework of space missions of the Italo-Dutch	BeppoSAX
satellite\footnote{\tt http://www.asdc.asi.it/bepposax/}
(Satellite per Astronomia X, and Beppo in the honor of Giuseppe
Occhialini) \cite{bepposax}, and NASA's Compton observatory, where
the BATSE experiment\footnote{\tt
http://www.batse.msfc.nasa.gov/batse/} (Burst and Transient Source
Experiment) \cite{batse} was performed, made it possible to study
the properties of the space distribution of gamma-ray bursts,
which, to a first approximation, are uniformly distributed in the
sky.

Among recent studies concerning this subject we distinguish the
analysis of the possible relation between the distribution of
gamma-ray bursts from the BATSE catalog and the large-scale
structure by Williams and Frey ~\cite{grb_apm}. The authors
analyze the effect of the local large-scale structure on the
apparent positions of gamma-ray bursts in the case of weak
lensing. This effect may show up for distant  ($z>4$) events as an
anticorrelation of the positions of gamma-ray bursts and galaxies
in clusters. The authors studying gamma-ray bursts report such an
anticorrelation for galaxies at $z \sim 0.2-0.3$, which they found
based on the optical magnitudes and positions of galaxies measured
using the APM (Automatic Plate Measuring Facility).

\begin{figure*}
\centerline{
\psfig{figure=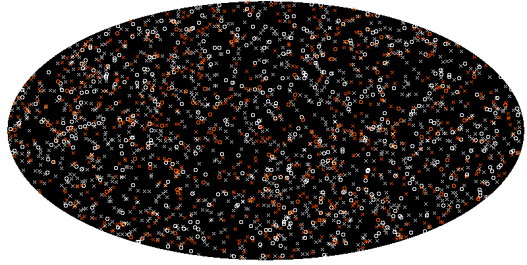,angle=-90,width=7cm}
}
\caption{Distribution of gamma-ray bursts on the sky. The gray
and white symbols (the circles and crosses correspond to short and
long bursts, respectively) show the  BeppoSAX and BATSE data,
respectively.  }
\label{f1}
\end{figure*}

\begin{figure*}
\centerline{\vbox{
\hbox{
\psfig{figure=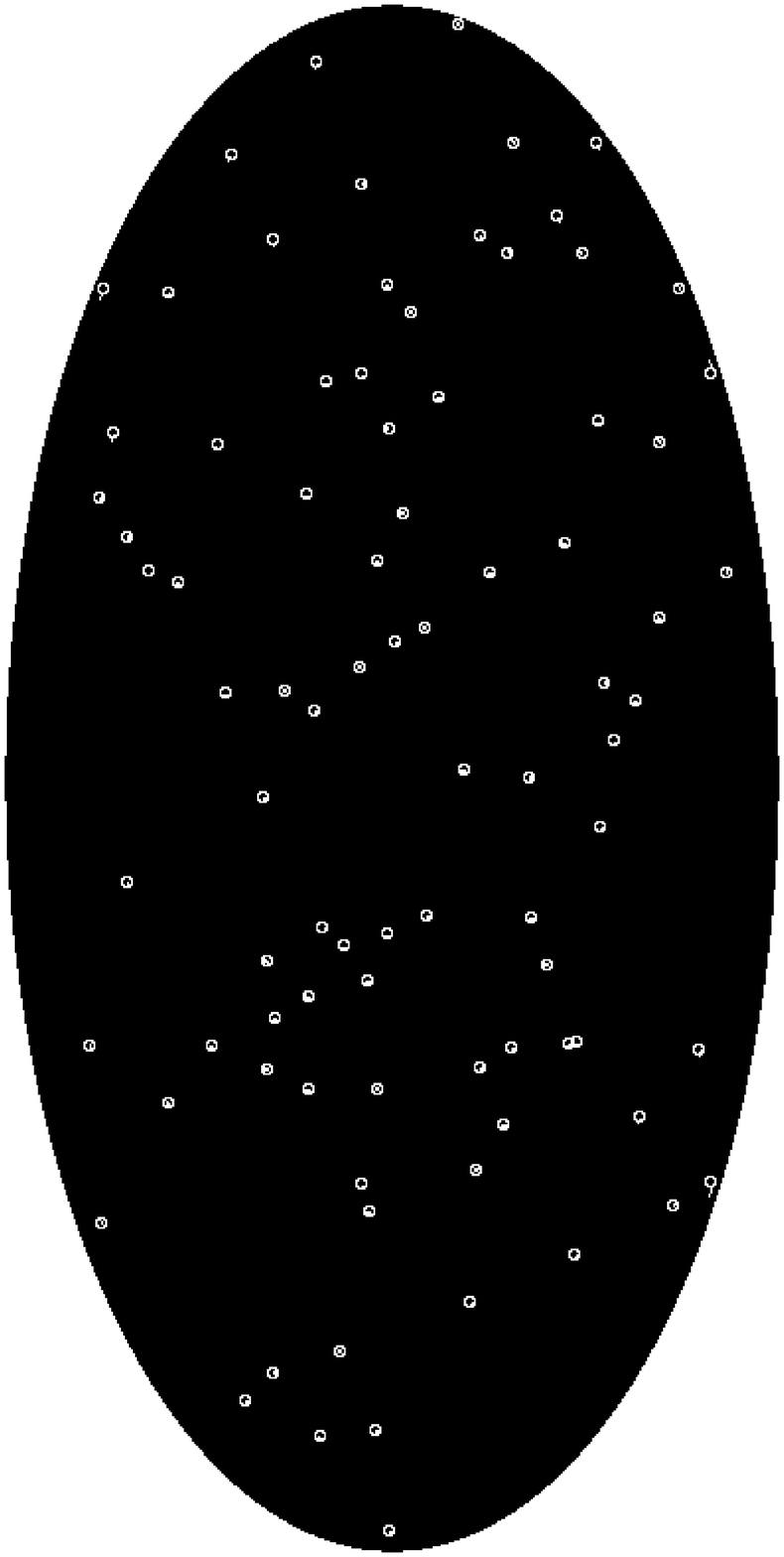,angle=-90,width=7cm}
\psfig{figure=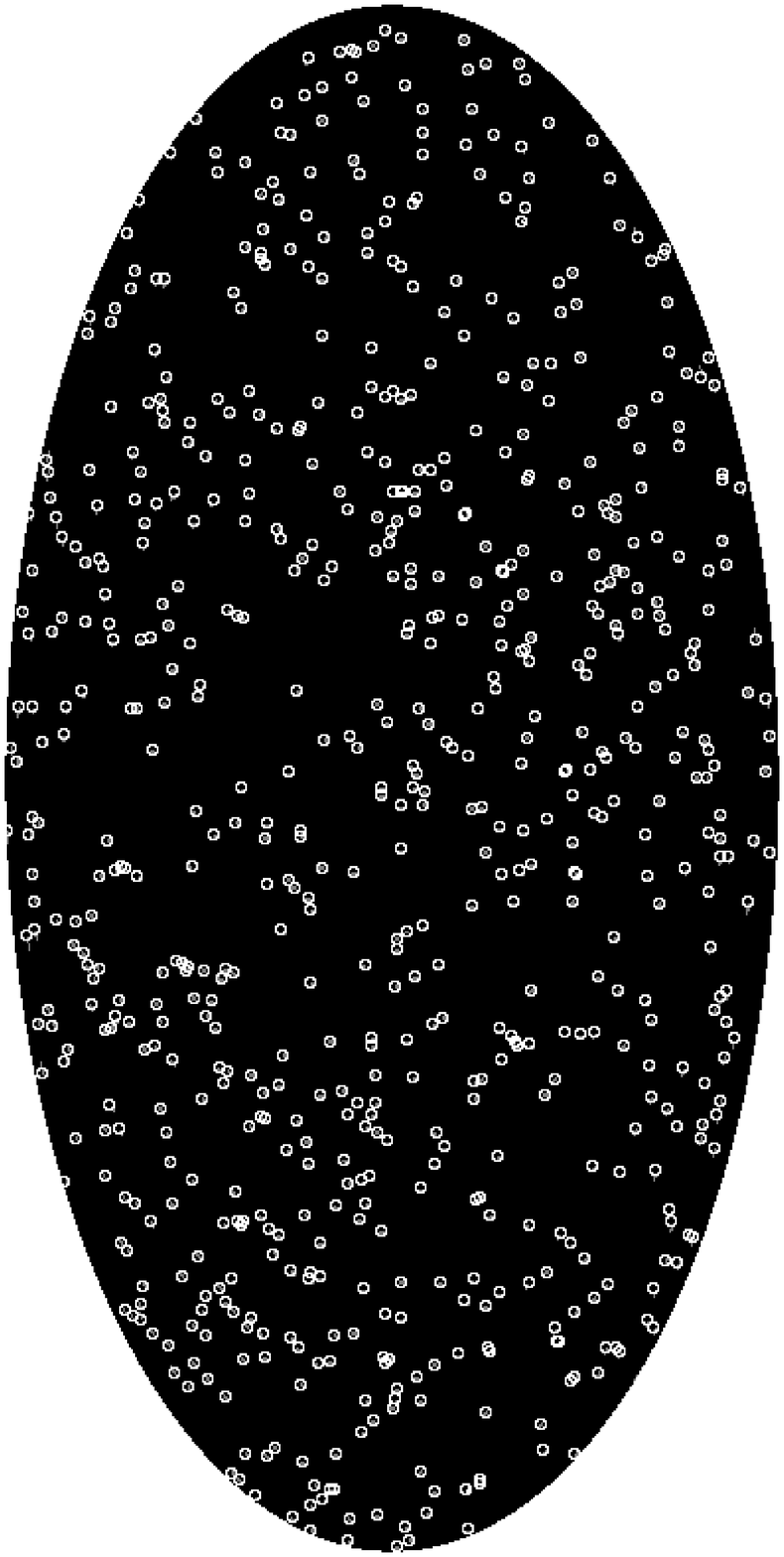,angle=-90,width=7cm}
} \hbox{
\psfig{figure=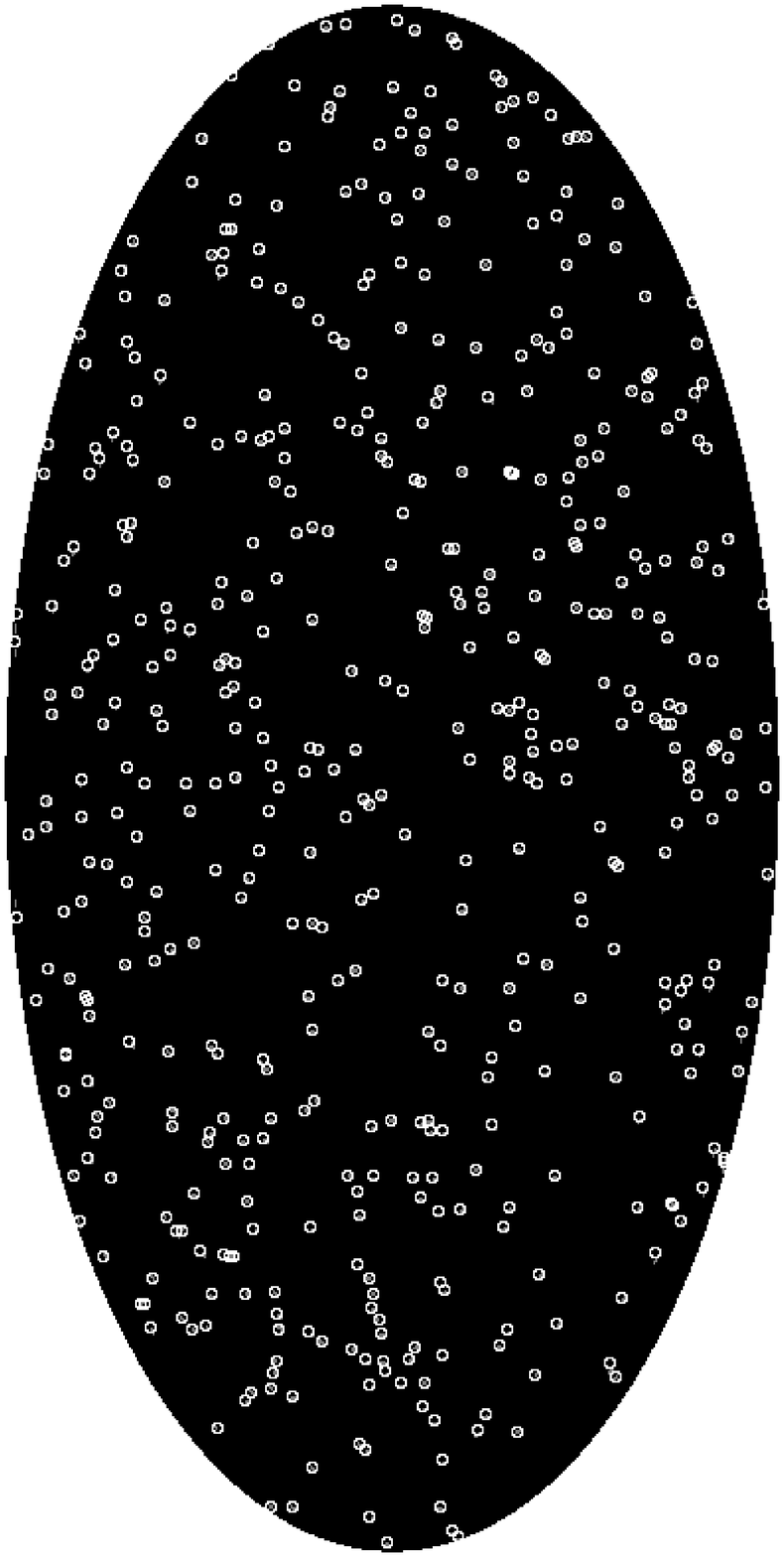,angle=-90,width=7cm}
\psfig{figure=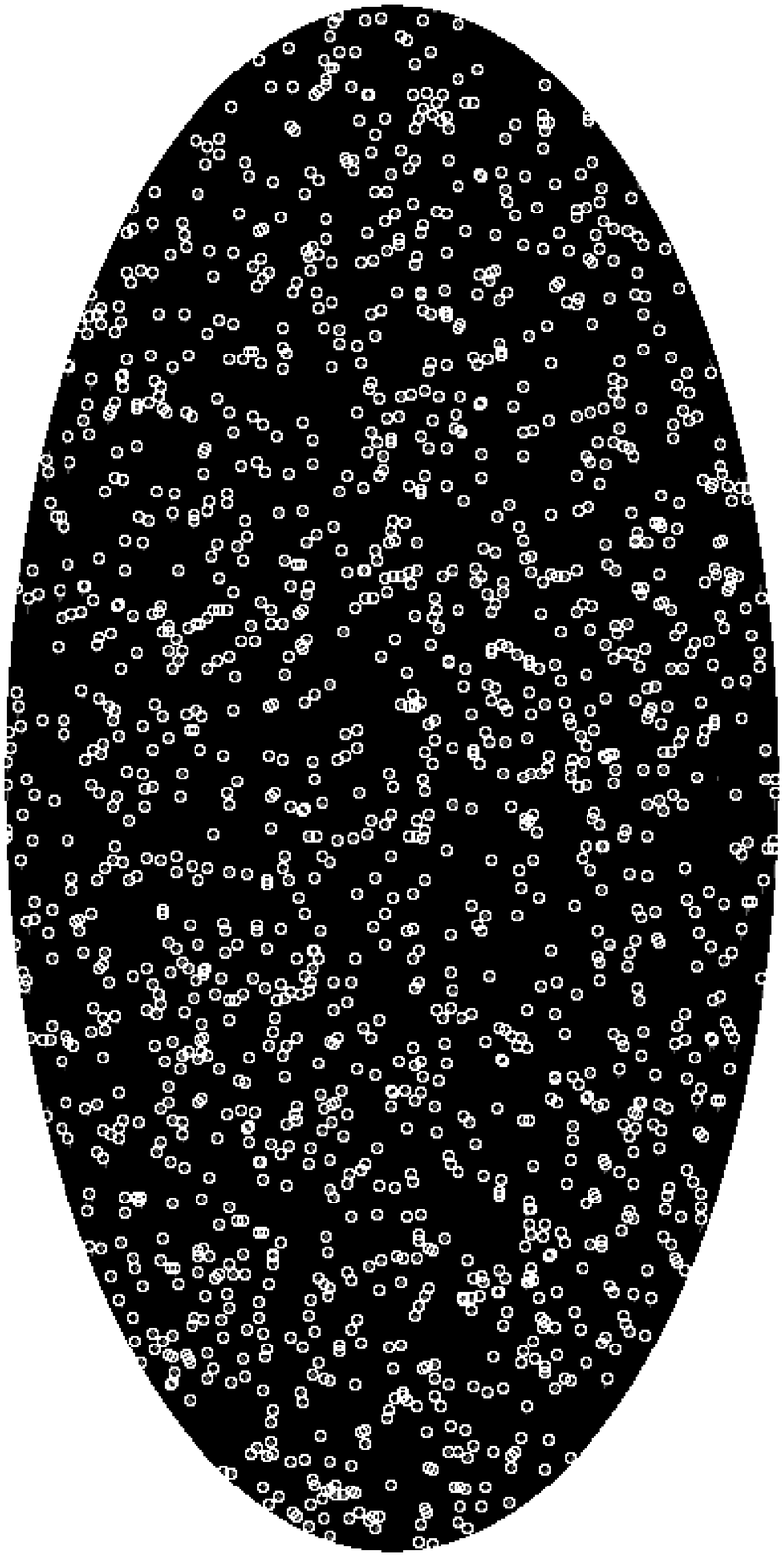,angle=-90,width=7cm}
}}}
 \caption{Distribution of the gamma-ray bursts on the sky. The top left panel shows the
BeppoSAX data for bursts with $t<2$\,s. The top right panel shows
the BeppoSAX data for bursts with $t>2$\,s. The bottom left panel
shows the BATSE data for $t<2$\,s. The bottom right panel shows
the BATSE data for $t>2$\,s.  }
\label{f2}
\end{figure*}

In another paper, M\'esz\'aros~et al. \cite{grb_vor} use various
methods (the Voronoi tesselation diagrams, the minimal spanning
tree, and multifractal spectra) to analyze the distribution of the
BATSE gamma-ray bursts. The above authors break the list of
objects into subsamples of sources with different signal durations
($t<2$\,s, \mbox{$2<t<10$\,s,} $t>10$\,s) and apply the above
methods to each subsample. They found the results for the first
two groups to deviate from the results obtained for uniformly
distributed simulated data. Based on the  data obtained the
authors \cite{grb_vor} discuss the validity of the cosmological
principle.

We use the  BeppoSAX (781 source, the energy interval covered:
0.1--200\,keV) and BATSE (2037 sources, 20\,keV--2\,MeV) catalogs
adopted from the sites of the corresponding experiments.
We subdivide each catalog into two subsamples containing short
(i.e., lasting $t<2$\,s) and long ($t>2$\,s) events. We thus
obtain four source subsamples, which we analyze separately using
the same procedure. Figure~1 shows the positions of all gamma-ray
bursts from the BeppoSAX and BATSE catalogs. Figure~2  shows the
positions of short and long BeppoSAX and BATSE gamma-ray bursts on
separate sky charts.

In this paper we analyze the statistical correlation properties of
the GRB distribution on the sky with respect to the distribution
of the cosmic microwave background (CMB). We proceed from the
assumption that gamma-ray bursts should be associated with massive
galaxies, and that the
positions of these galaxies are correlated with the large-scale
structure. In this case there is hope that we may find deviations
in the sky distribution of CMB fluctuations in the areas where the
bursts are projected. Note that the large-scale structure may show
up in CMB fluctuations via the Zel'dovich--Sunyaev effect and the
integrated Sachs--Wolfe effect~\cite{swe}. This former effect
manifests itself via the interaction of CMB photons with hot
electrons of galaxy clusters \cite{zse} and is observed on the
angular scales of about 4\arcmin\, or smaller. The latter effects
occurs when the photons move in a gravitational field with a
variable potential that arises during
the formation of the large-scale structure and expansion of the
Universe, and should be observed on the angular scales greater
than 200\arcmin.

To study the correlation properties of the positions of gamma-ray
bursts and CMB peaks, we use the CMB map based on the results of
the experiment that has been conducted for seven years within the
framework of the WMAP\footnote{{\tt http://lambda.gsfc.nasa.gov}}
(Wilkinson Microwave Anisotropy Probe) experiment aimed to
integrate the signal from the entire sky \cite{wmap7ytem}. The CMB
signal was reconstructed from multifrequency observations using
the method of internal linear combination (ILC) of the background
components~\cite{wmapresults}. This procedure yielded the CMB map,
which is also referred to as the ILC map, and which is used to
analyze low-order harmonics with multipole numbers $\ell\le150$.
The ILC map is based on observations made in five channels:
23\,GHz (the K band), 33\,GHz (the Ka band), 41\,GHz  (the Q
band), 61\,GHz (the V band), and 94\,GHz (the W band).

Given that one of the central problems in statistical studies of
gamma-ray bursts is due to the large error box size, about
1\degr$\times$1\degr, of the available source positions, we
operate with pixels of about the same or even greater size in
order to avoid uncertainties in the analysis of the sky
distribution of the events.

The methods we use in this paper and our results are laid out in
the following way. First we describe the methods that we use to
pixelize the distribution of bursts and to correlate the maps. We
then analyze the statistics of CMB deviations in the gamma-ray
burst areas. At the next stage, we compare the BATSE--BeppoSAX and
BATSE--CMB correlation maps. We finally discuss the results in the
concluding section.

\section{MOSAIC CORRELATION METHOD}

To analyze the map properties on different angular scales, we
expand the signal distributed on the sphere into the spherical
harmonics (multipoles):
\begin{equation}
\Delta S(\theta,\phi)= \sum_{\ell=1}^{\infty}\sum_{m=-\ell}^{m=\ell}
       a_{\ell m} Y_{\ell m} (\theta, \phi)\,.
\label{eq1}
\end{equation}
Here $\Delta S(\theta,\phi)$ are the  variations of signal on the
sphere in polar coordinates; $\ell$ is the number of the
multipole, and $m$ is the number of the mode of the multipole. The
spherical harmonics are determined as
\begin{eqnarray}
Y_{\ell m}(\theta,\phi) = \sqrt{{(2\ell+1)\over 4\pi}{(\ell-m)!
\over (\ell+m)!}}P_\ell^m(x) e^{i m\phi}, \\
 ~x=\cos\theta\,,\nonumber \label{eq2}
\end{eqnarray}
where $P_\ell^m(x)$ are the associated Legendre polynomials. The
expansion coefficients $a_{\ell m}$ for a continuous function
$\Delta S(x,\phi)$ can be written as:
\begin{equation}
a_{\ell m}=\int^1_{-1}dx\int^{2\pi}_0 \Delta S(x,\phi)
Y^{*}_{\ell m}(x,\phi) d\phi\,,
\label{eq3}
\end{equation}
where $Y^{*}_{\ell m}$ denotes the complex conjugate of	 $Y_{\ell m}$.

The correlation properties of two maps of the signal distribution
on the sphere can be described, on a given angular scale, by a
correlation coefficient for the corresponding multipole $\ell$ as:
\begin{equation}
K(\ell) = \frac{1}{2} \frac{\sum\limits_{m=-\ell}^\ell t_{\ell
m}s^*_{\ell m} + t^*_{\ell m}s_{\ell m}}
       {(\sum\limits_{m=-\ell}^\ell |t_{\ell m}|^2
     \sum\limits_{m=-\ell}^\ell |s_{\ell m}|^2)^{1/2}}\,,
\label{eq4}
\end{equation}
where $t_{\ell m}$ and $s_{\ell m}$ are the variations of two
signals in a harmonic representation. The coefficient $K(\ell)$
can be used to assess the correlation between the harmonics on the
sphere, i.e., to compare the properties of maps on a given angular
scale. However, in the case of a search for correlated areas,
which do not repeat in other regions of the sphere, this approach
smears such single areas in the process of averaging over the
sphere within a certain harmonic. In this case it becomes
practically impossible to identify the correlated areas.

Verkhodanov et al.~\cite{cormap} proposed an approach, which was
implemented in the second release of the GLESP package
\cite{glesp2} (the {\tt difmap} utility). The proposed procedure
makes it possible to find correlations between two maps in the
areas of a certain angular size. In this method, each pixel with
number $p$ subtending the solid angle  $\Xi_p$ is assigned the
cross-correlation coefficient between the data of the two maps on
the corresponding area. Thus a correlation map is constructed for
two signals  $T$ and $S$, where the value of each pixel $p$
($p=1,2,...,N_0$, \mbox{and $N_0$ is the} total number of pixels
on the sphere) with the subtending angle $\Xi_p$, and computed for
the sphere maps with the initial resolution determined by
$\ell_{max}$ is \mbox{equal to}
\begin{eqnarray}
\nonumber   K(\Xi_p|\ell_{max}) = \hspace{3cm} \\
 \frac{\sum\sum\limits_{p_{ij}\in\Xi_p}
	   (T(\theta_i,\phi_j) - \overline{T(\Xi_p))}
	   (S(\theta_i,\phi_j) - \overline{S(\Xi_p))}}
	{\sigma_{T_p}\sigma_{S_p}}\,.
\label{eq5}
\end{eqnarray}
Here $T(\theta_i,\phi_j)$ is the value of the signal $T$ in the
pixel with the coordinates $(\theta_i,\phi_j)$ for the initial
resolution of the pixelization of the sphere; $S(\theta_i,\phi_j)$
is the value of the other signal in the same area;
$\overline{T(\Xi_p)}$ and \mbox{$\overline{S(\Xi_p)}$ } are the
mean values averaged over the area $\Xi_p$ and obtained from the
data of the maps with higher resolution determined by
$\ell_{max}$, and  $\sigma_{T_p}$ and $\sigma_{S_p}$ are the
corresponding standard deviations in the area considered.

\section{CMB SIGNAL STATISTICS IN THE AREAS OF GAMMA-RAY BURSTS}

\begin{figure*}
\centerline{\vbox{
\hbox{
\psfig{figure=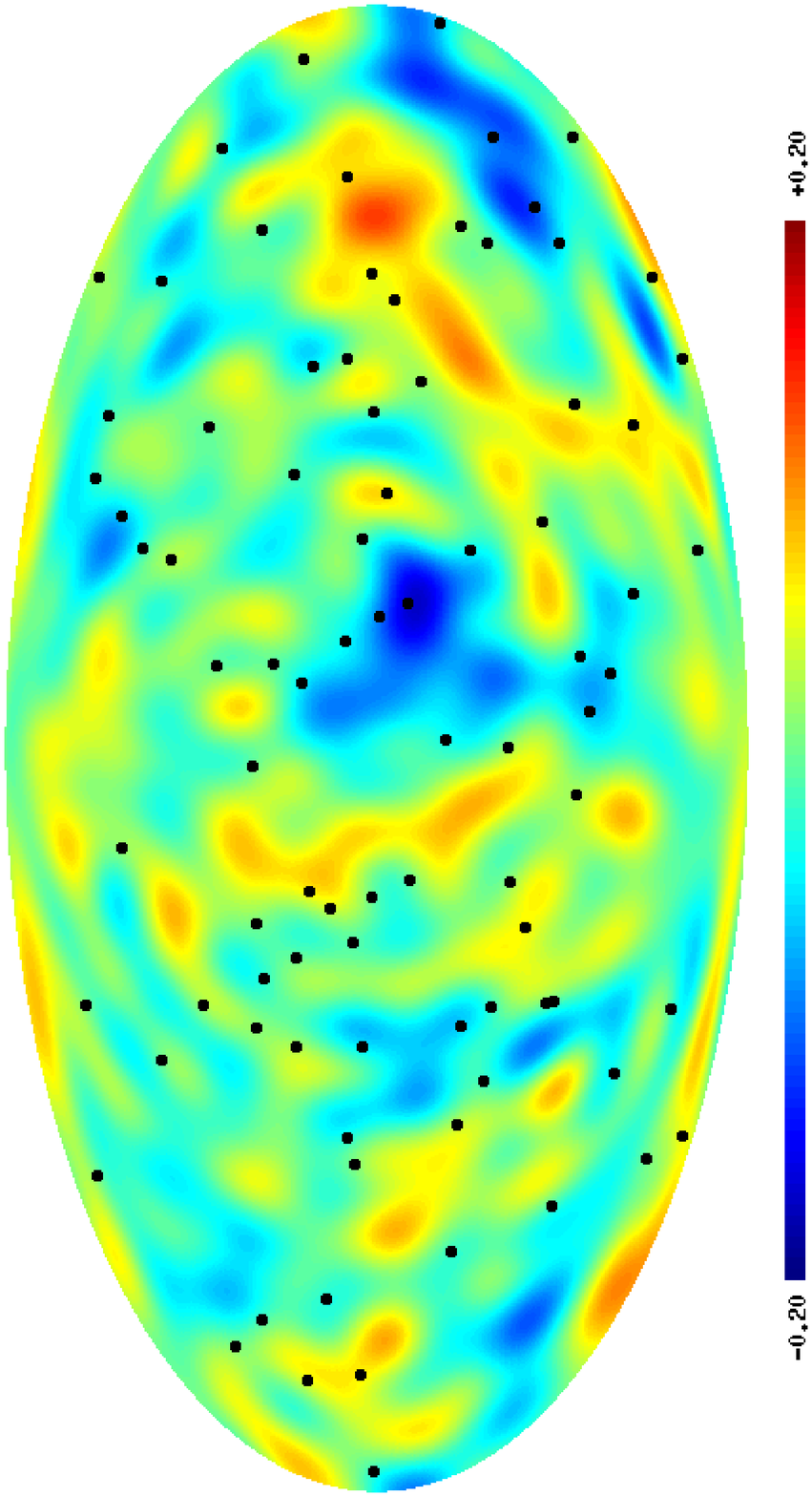,angle=-90,width=7cm}
\psfig{figure=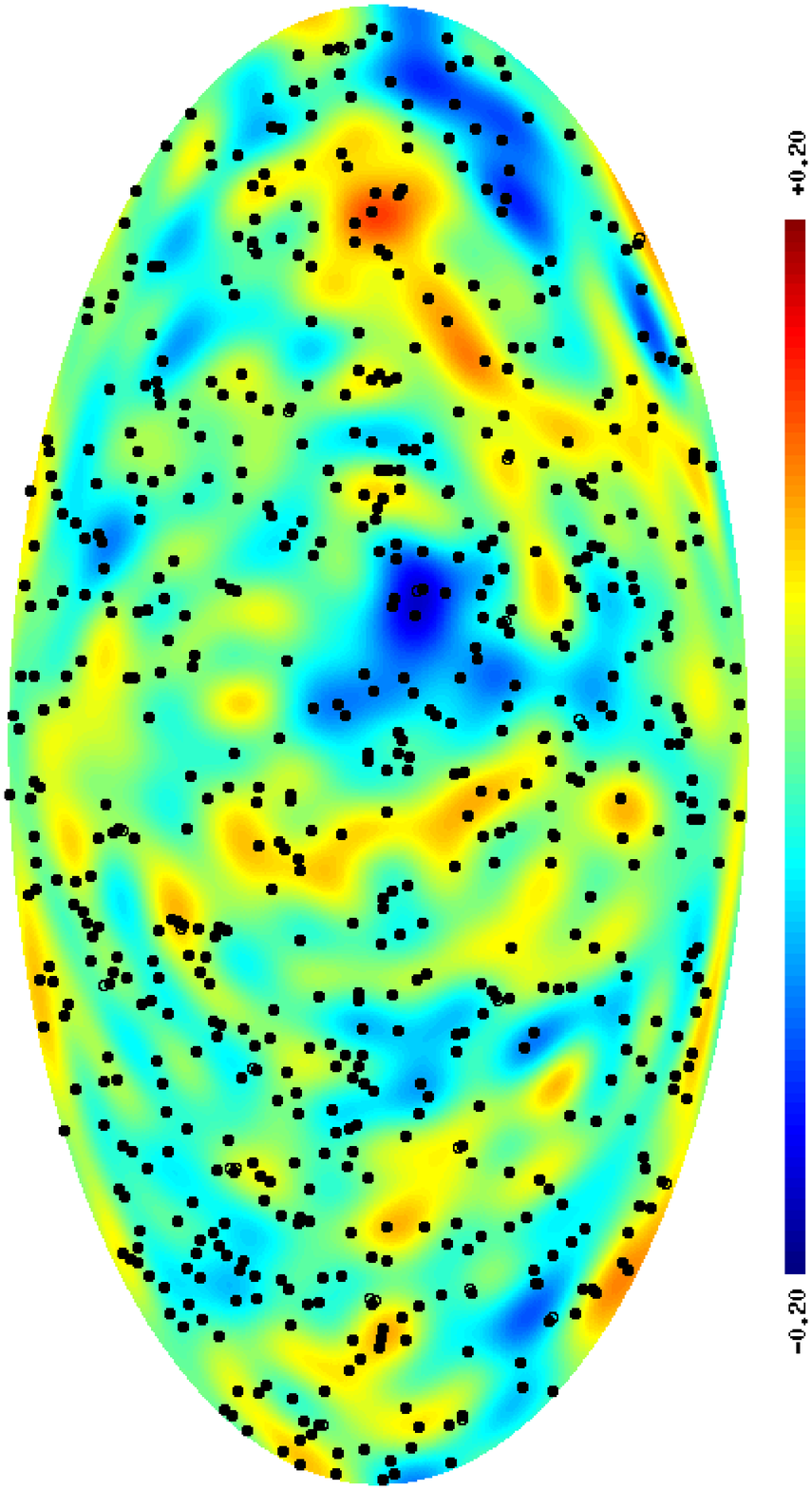,angle=-90,width=7cm}
} \hbox{
\psfig{figure=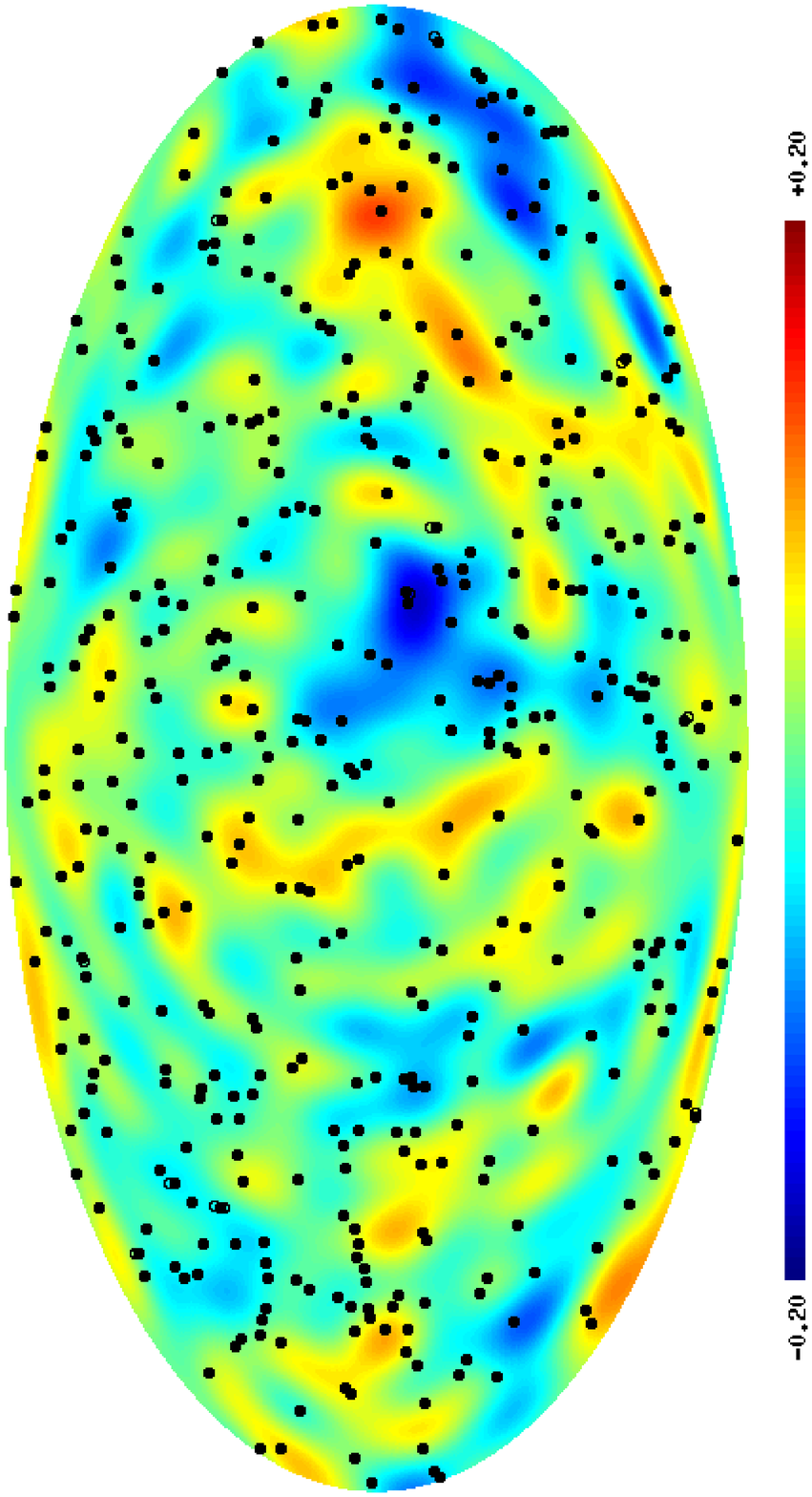,angle=-90,width=7cm}
\psfig{figure=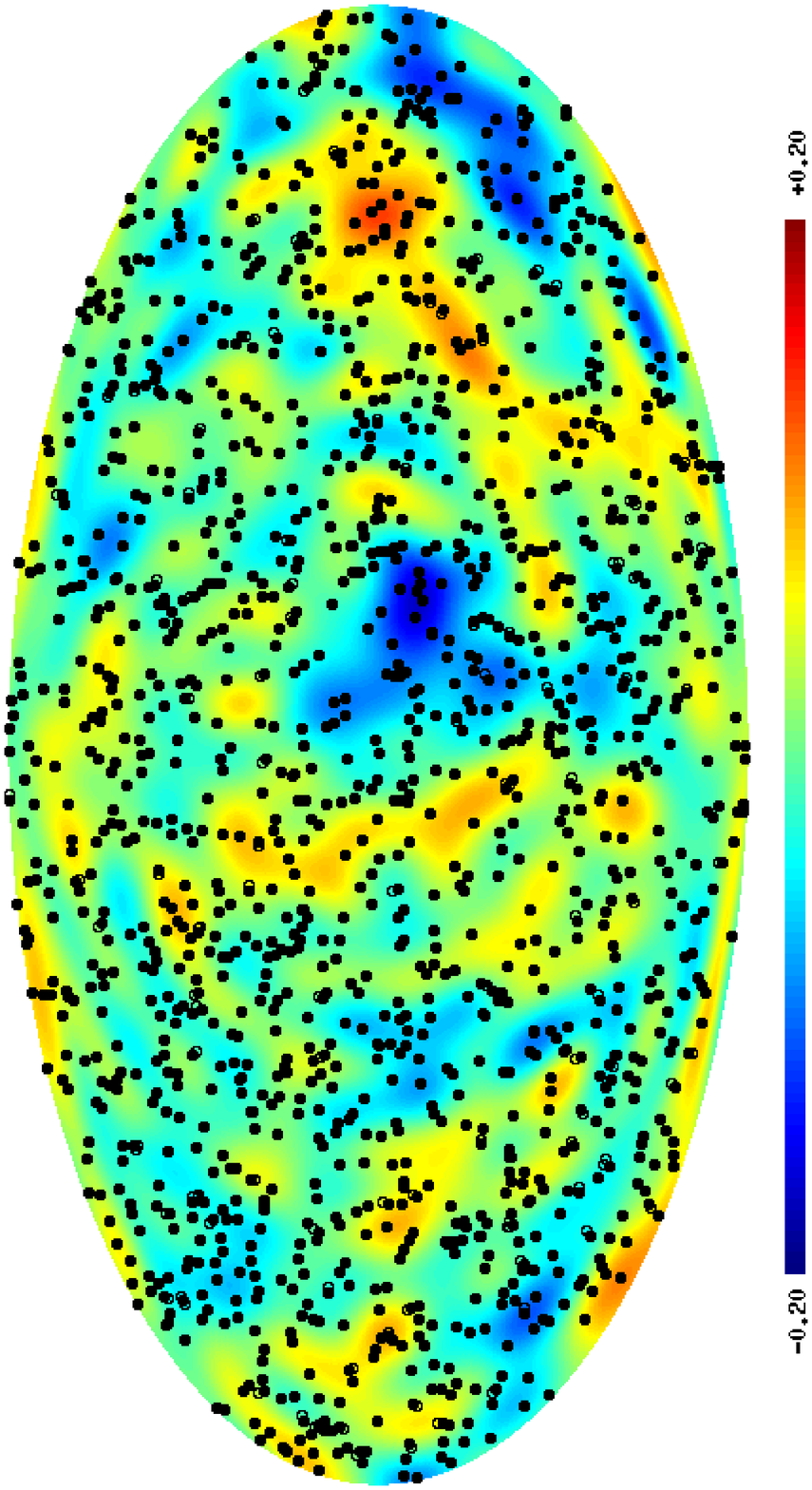,angle=-90,width=7cm}
}}}
 \caption{Positions of gamma-ray bursts from different subsamples on the CMB maps
with a resolution of $\ell_{max}=20$. The top left and top right
panels show the BeppoSAX data for bursts with  $t<2$\,s and
$t>2$\,s, respectively. The bottom left and bottom right panels
show the data for the BATSE bursts  with  $t<2$\,s and	$t>2$\,s,
respectively.}
\label{f3}
\end{figure*}

\begin{figure*}
\centerline{\vbox{
\hbox{
\psfig{figure=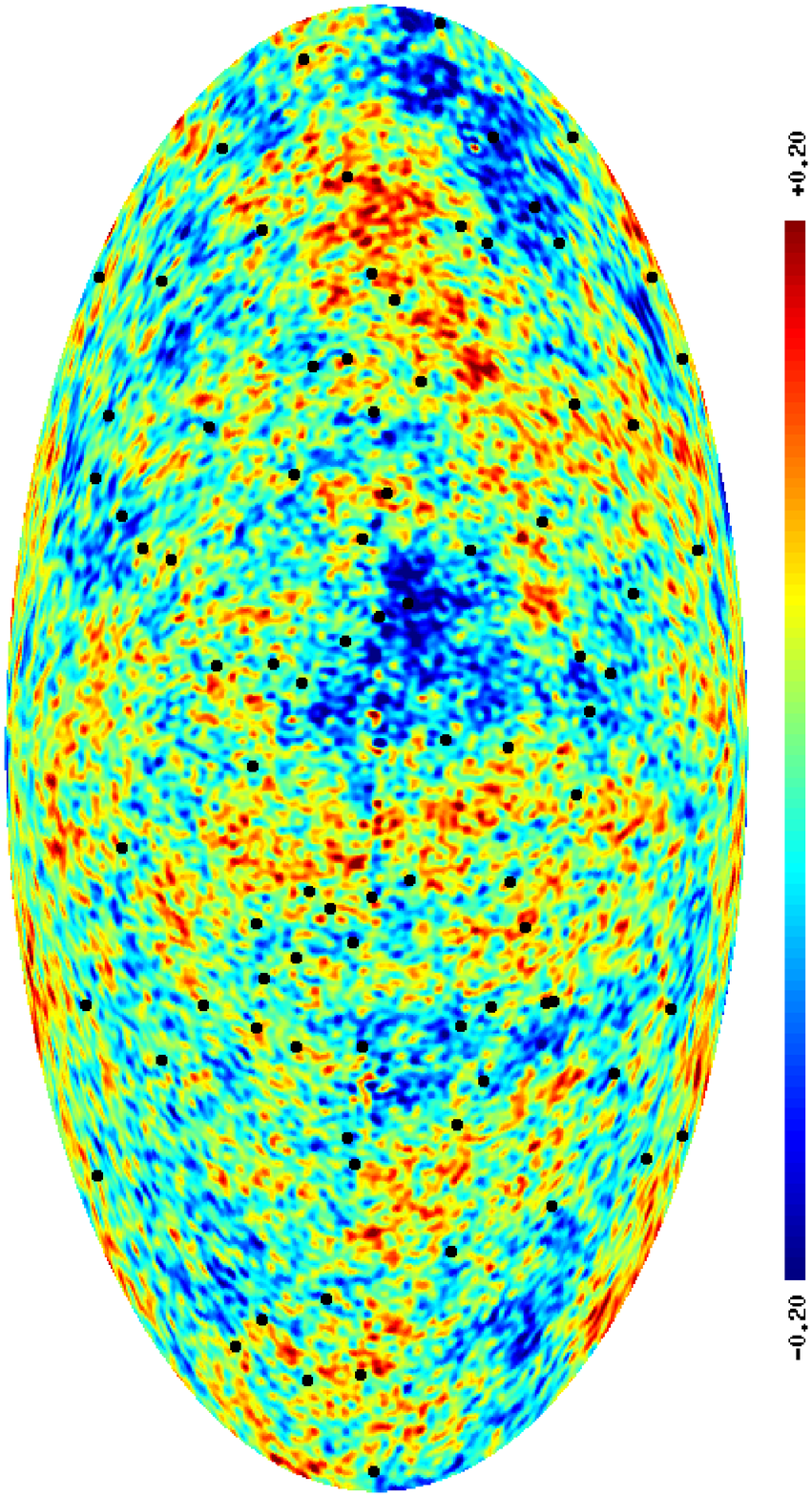,angle=-90,width=7cm}
\psfig{figure=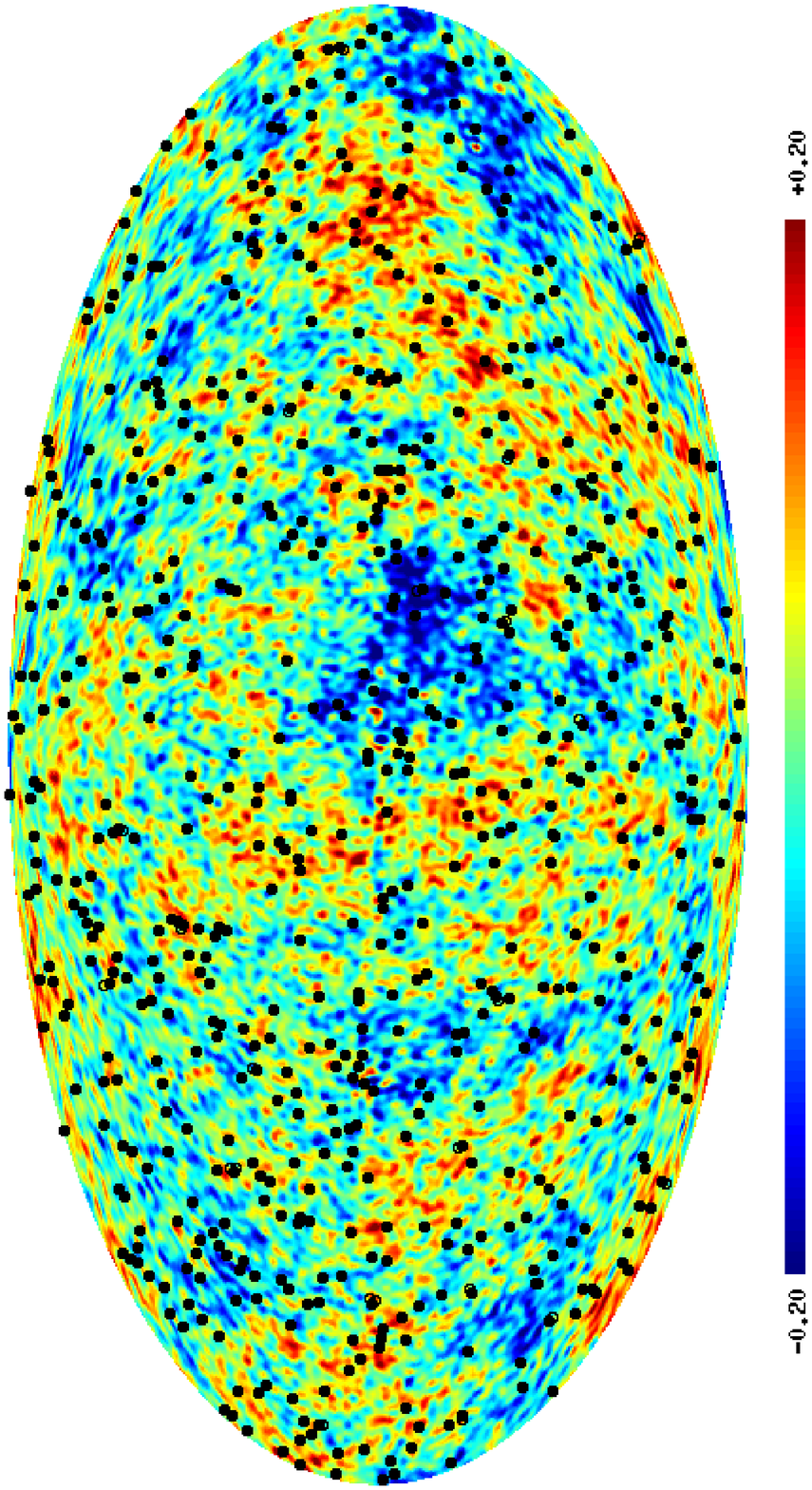,angle=-90,width=7cm}
} \hbox{
\psfig{figure=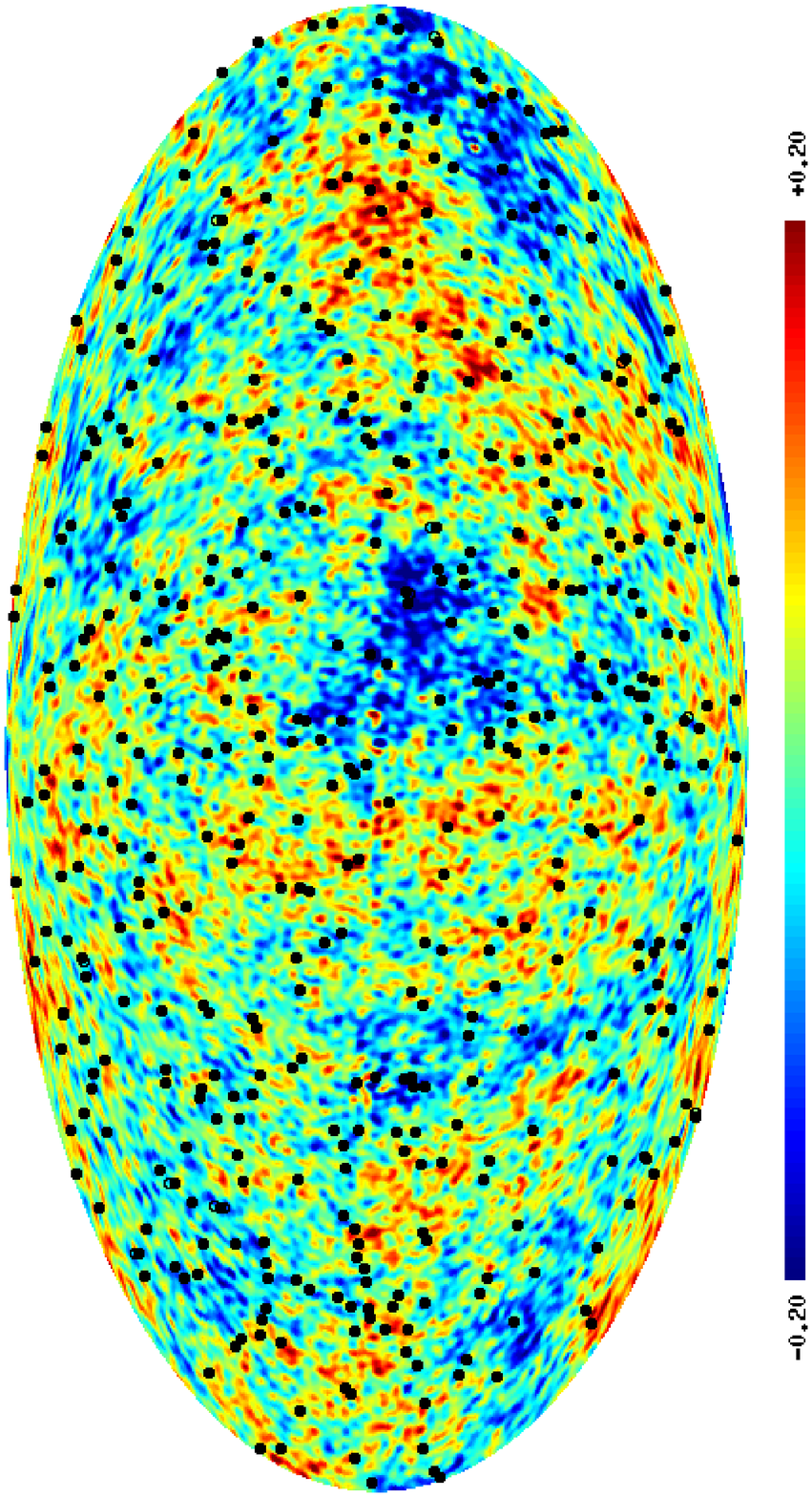,angle=-90,width=7cm}
\psfig{figure=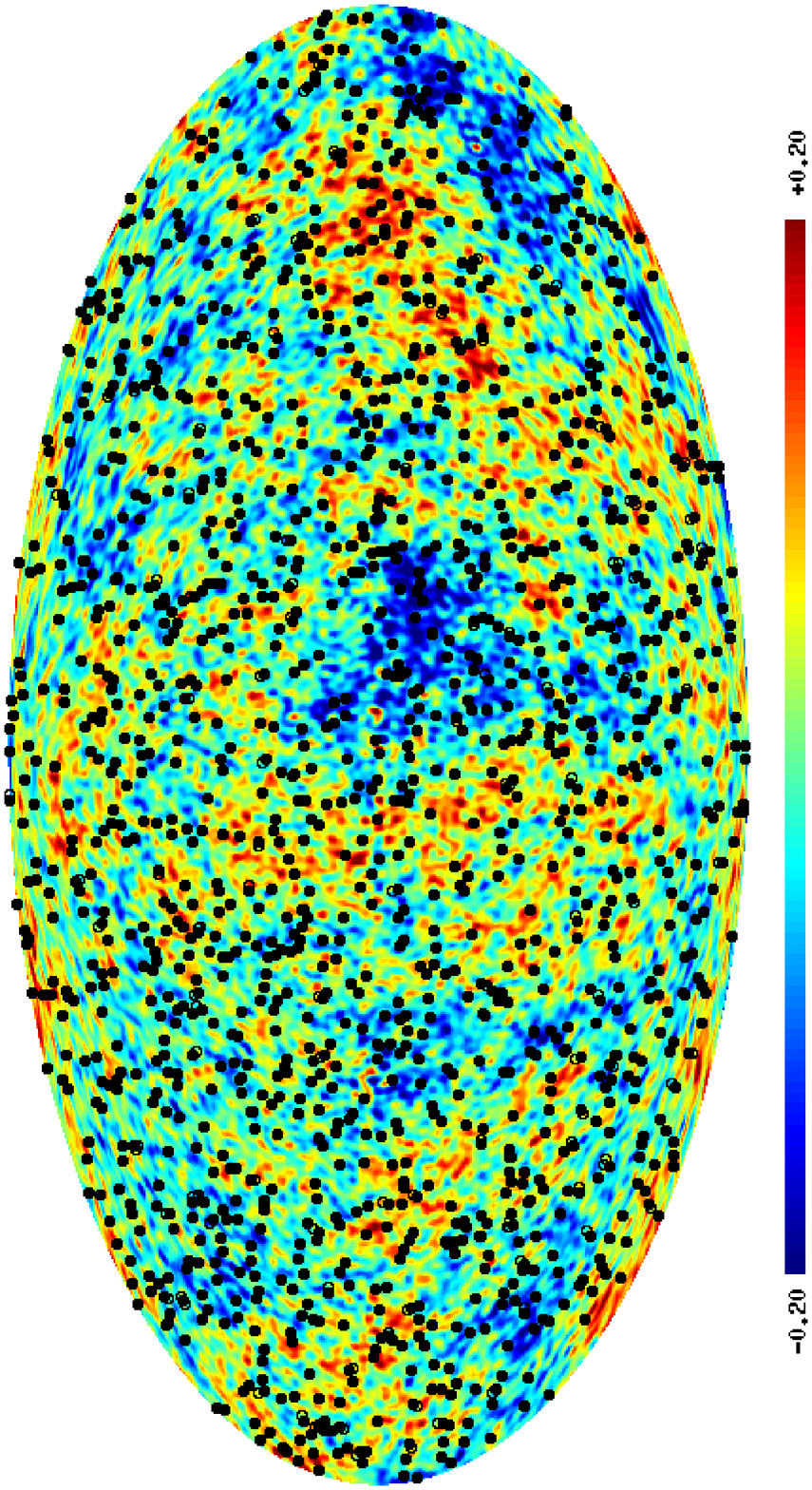,angle=-90,width=7cm}
 }}}
\caption{Positions of gamma-ray bursts from different subsamples
on the CMB maps with a resolution of $\ell_{max}=150$. The top
left and top right panels show the BeppoSAX data for bursts with
$t<2$\,s and  $t>2$\,s, respectively. The bottom left and bottom
right panels show the data for the BATSE bursts	 with  $t<2$\,s
and $t>2$\,s, respectively.  }
\label{f4}
\end{figure*}

In the first method  the statistical properties of distributions
of GRB positions versus the CMB signal are tested by simply
analyzing the pixel value measurements on the CMB maps. We
performed pixel measurements using the {\tt mapcut} procedure of
the GLESP package. Figures~3 and 4 show the positions of gamma-ray
bursts of the BeppoSAX and BATSE catalog subsamples on the CMB
maps with the resolution of  260\arcmin\, \mbox{($\ell_{max}=20$)}
and 36\arcmin\, ($\ell_{max}=150$). We chose the resolution of
these maps to match the expected angular scale of the Sachs--Wolfe
effect and the limiting resolution of the WMAP mission.

To search for the eventual correlations, we computed the number of
GRB positions that fall onto the CMB pixels with negative signal
fluctuations (which may be due to the effects described above) in
the CMB maps of different resolution. The Table lists the
statistics of CMB pixel values in the gamma-ray burst areas for
the long- and short-event BATSE and BeppoSAX subsamples. It
includes the total number of sources in subsamples; the number of
sources located in CMB pixels with negative fluctuation values;
the expected number of pixels with negative CMB amplitudes
according to the results of 200 random Gaussian CMB signal
simulations modeled in terms of the $\Lambda$CDM cosmology, and
the 1$\sigma$-scatter of these quantities.

\begin{table}
\caption{{\bf Table.}~Statistics of the
CMB pixel values in the GRB areas for the  BATSE and BeppoSAX
subsamples. The columns are: $t$, the duration (s); mission name;
resolution of the CMB map (the number of the multipole); total
number of gamma-ray sources in the subsample ($N_t$); the number
of sources ($N_e$) located in the CMB pixels with negative
fluctuation values; expected number of pixels with negative CMB
amplitudes according to the results of 200 random Gaussian CMB
signal simulations, modelled in terms of the $\Lambda$CDM
cosmology, and the 1$\sigma$-scatter of these quantities}
\begin{tabular}{clrrrc}
\hline
  $t$, s &  Mission &$\ell_{max}$ & $N_t$ & $N_e$ & Model   \\
\hline
$<$2  &	 BATSE	  & 150	  &  497  &  244  &  249$\pm$11	 \\
$>$2  &	 BATSE	  & 150	  & 1540  &  763  &  769$\pm$19	 \\
$<$2  &	 BATSE	  &  20	  &  497  &  250  &  248$\pm$13	 \\
$>$2  &	 BATSE	  &  20	  & 1540  &  781  &  768$\pm$32	 \\
\hline
$<$2  &	 BeppoSAX & 150	  &   87  &   43  &   44$\pm$5	 \\
$>$2  &	 BeppoSAX & 150	  &  694  &  339  &  347$\pm$15	 \\
$<$2  &	 BeppoSAX &  20	  &   87  &   50  &   44$\pm$5	 \\
$>$2  &	 BeppoSAX &  20	  &  694  &  346  &  348$\pm$18	 \\
\hline
\end{tabular}
\end{table}

\begin{figure*}
\centerline{\vbox{
\hbox{
\psfig{figure=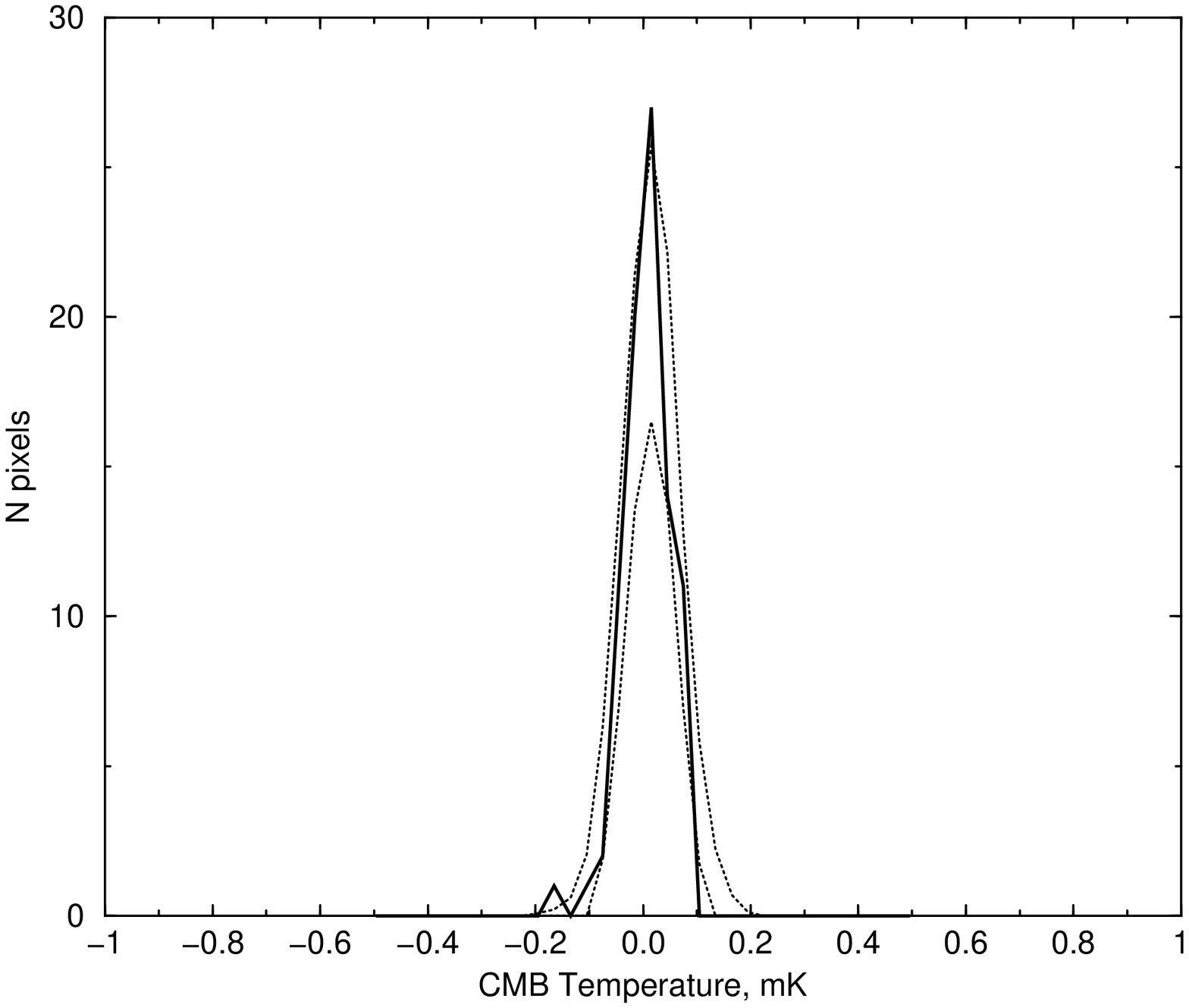,angle=-90,width=7cm}
\psfig{figure=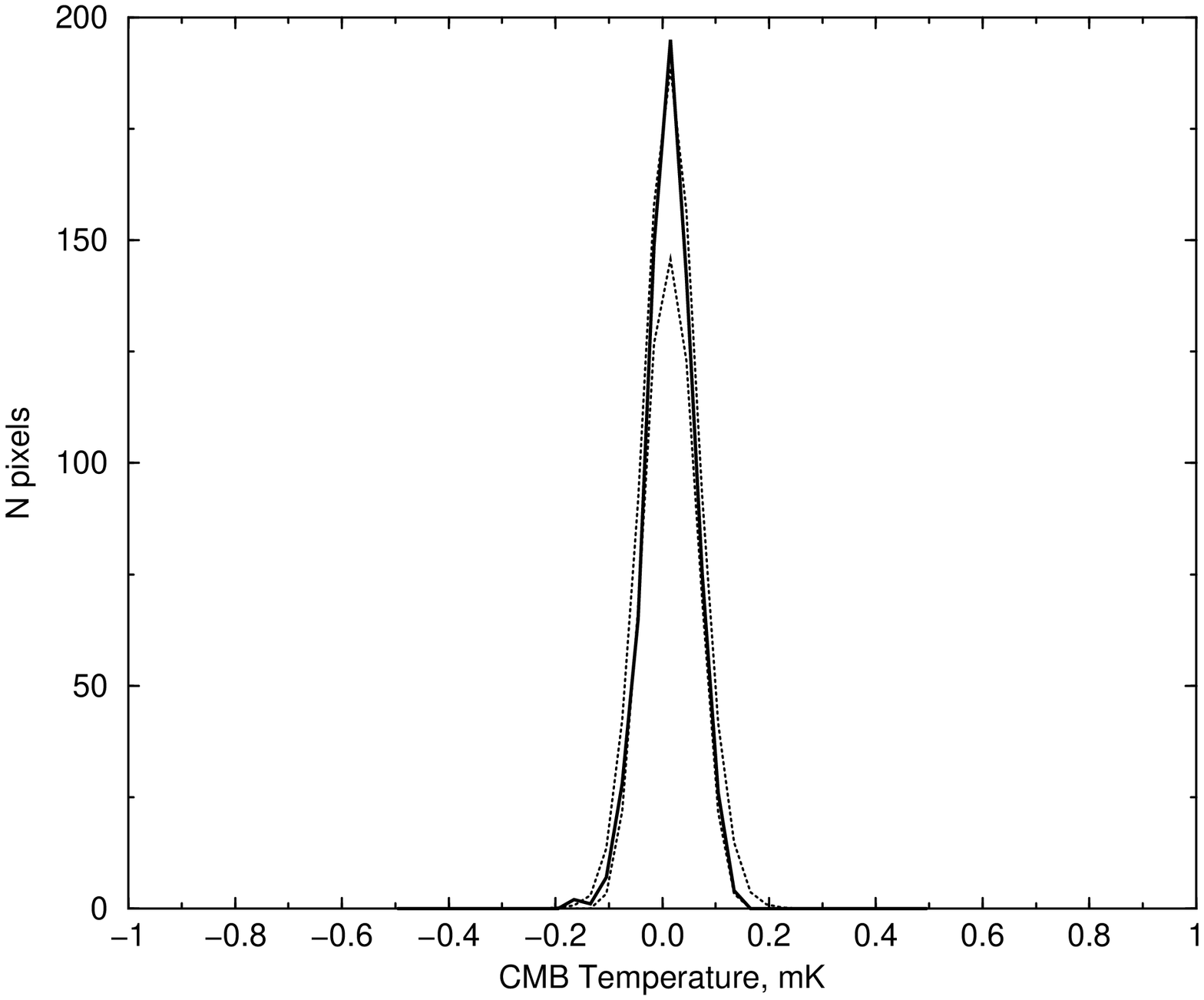,angle=-90,width=7cm}
} \hbox{
\psfig{figure=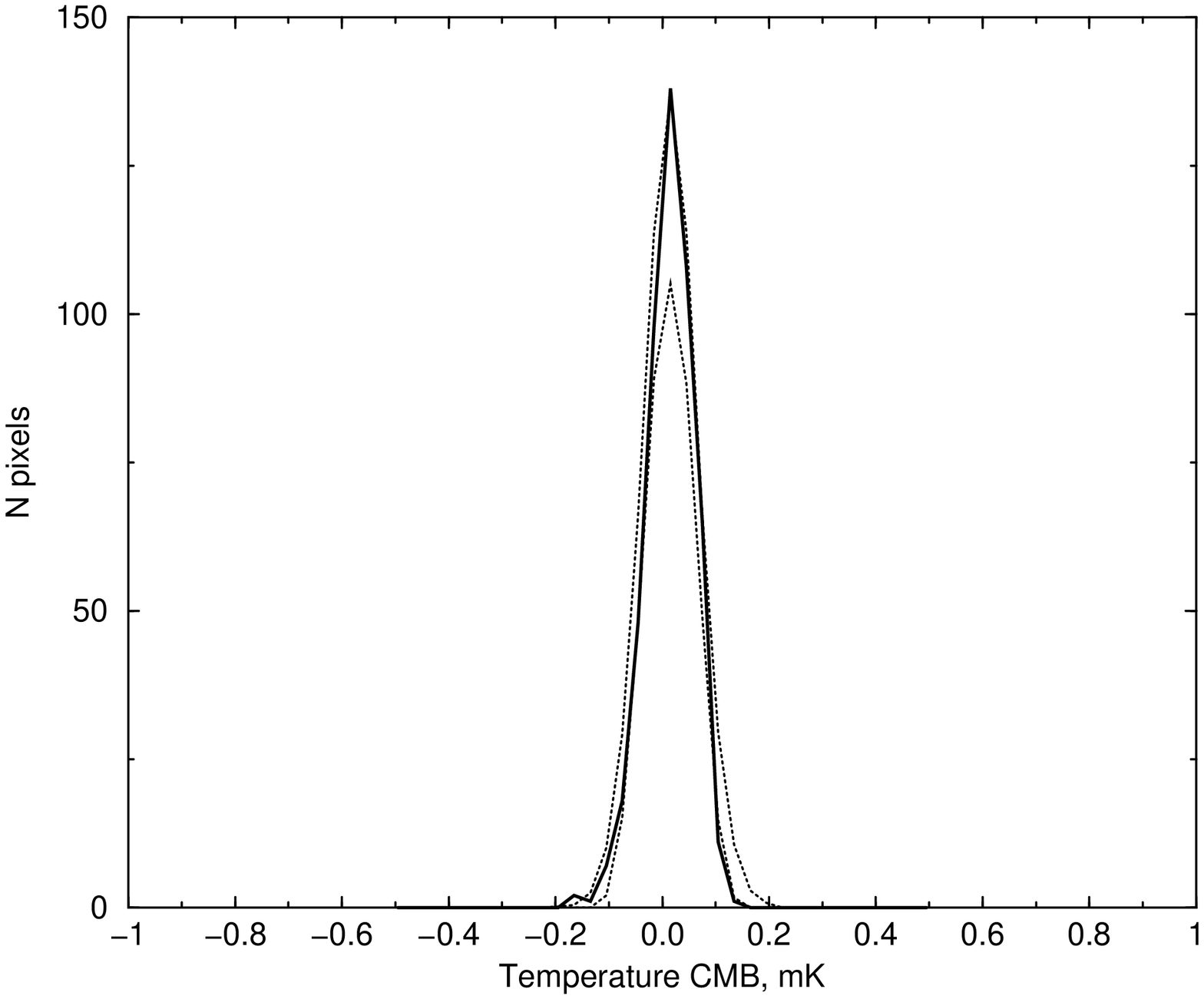,angle=-90,width=7cm}
\psfig{figure=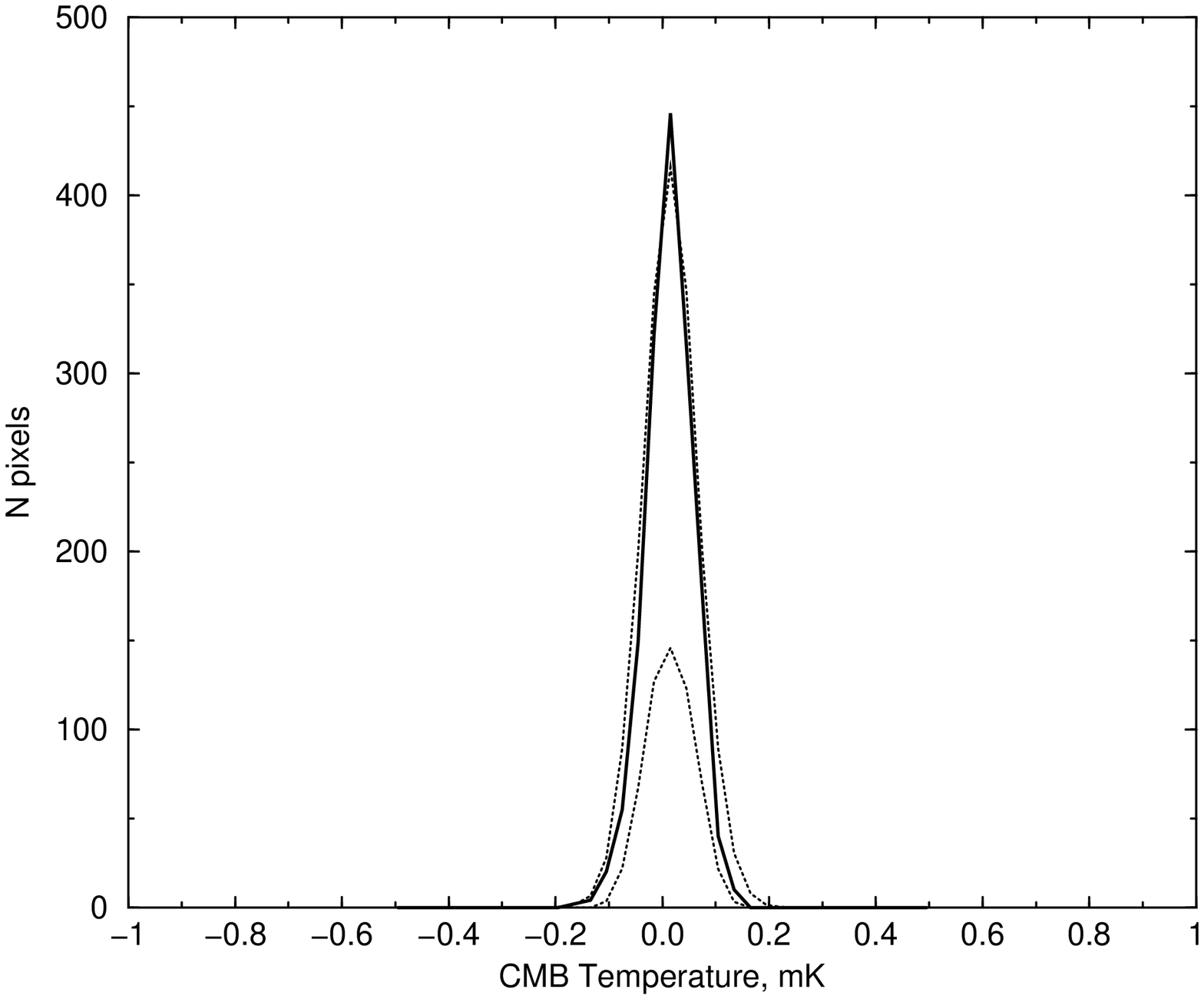,angle=-90,width=7cm}
 } }}
\caption{ Distribution of CMB fluctuations in the WMAP pixels
corresponding to the GRB positions on the maps with a resolution
of $\ell_{max}=20$. The top left and top right panels show the
BeppoSAX data for the bursts with  $t<2$\,s and	 $t>2$\,s,
respectively. The bottom left and bottom right panels show the
data for the BATSE bursts  with	 $t<2$\,s and  $t>2$\,s,
respectively. The dashed lines show the 1$\sigma$ scatter of CMB
values in the $\Lambda$CDM cosmological model.}
\label{f5}
\end{figure*}

\begin{figure*}
\centerline{\vbox{
\hbox{
\psfig{figure=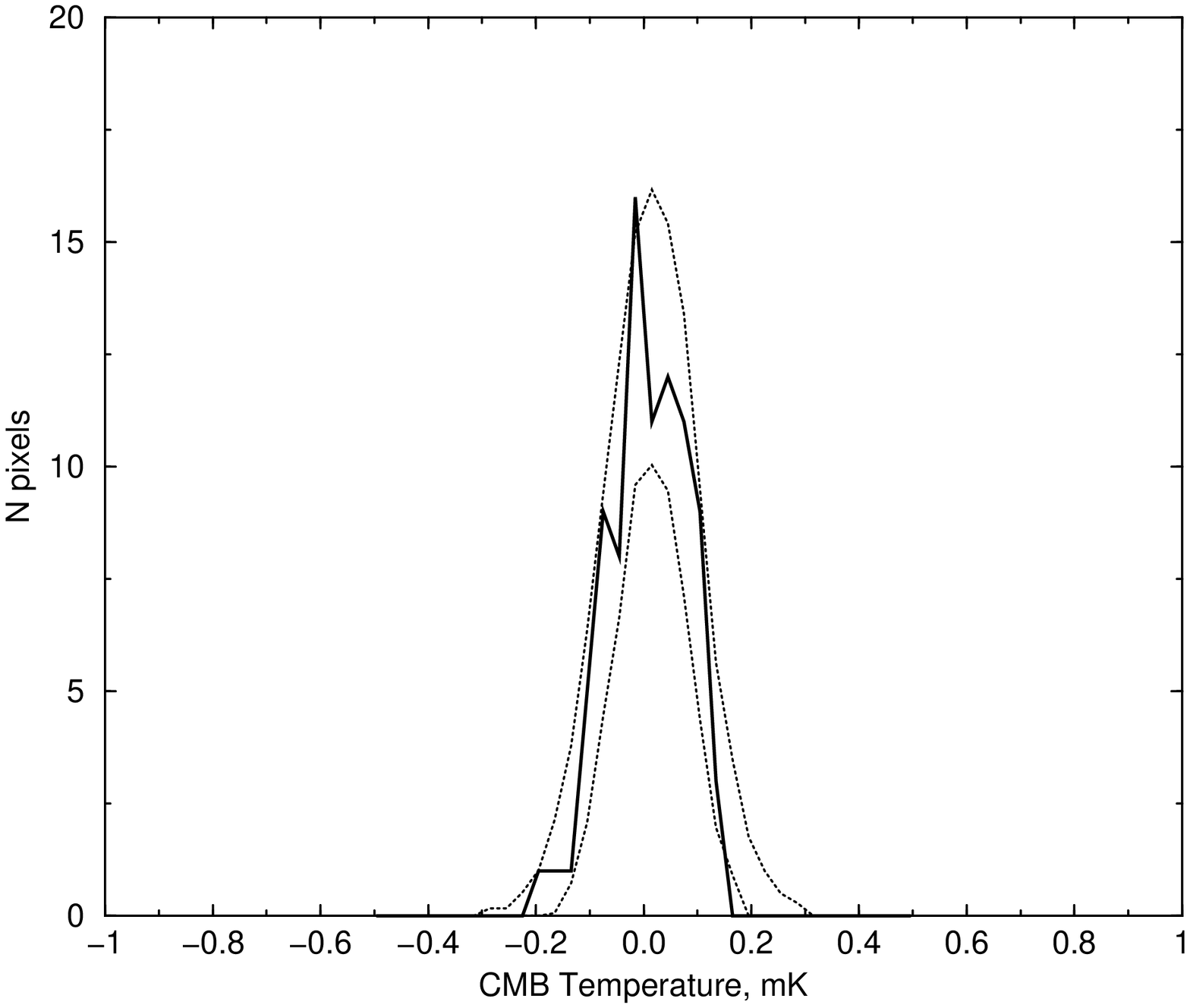,angle=-90,width=7cm}
\psfig{figure=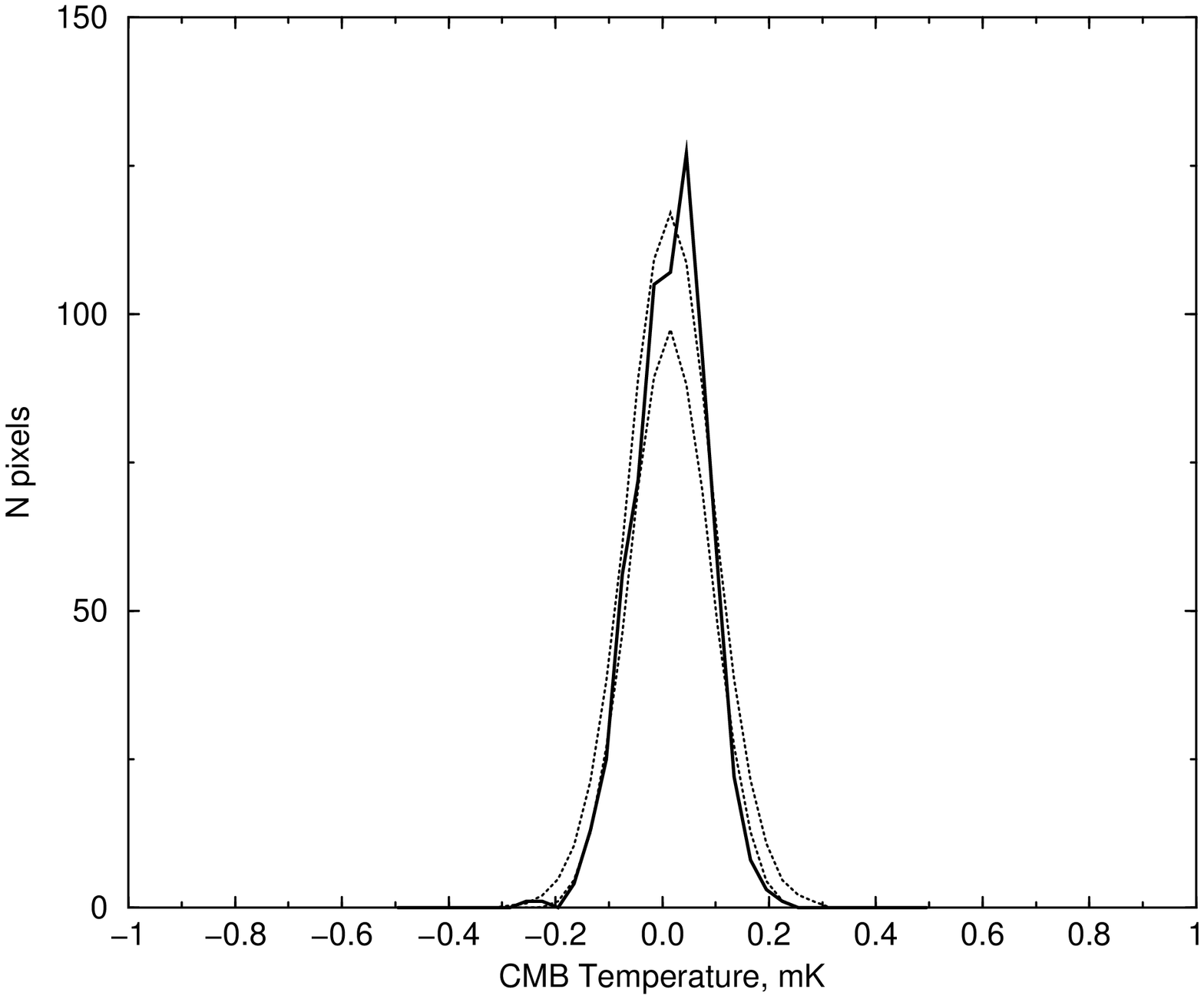,angle=-90,width=7cm}
} \hbox{
\psfig{figure=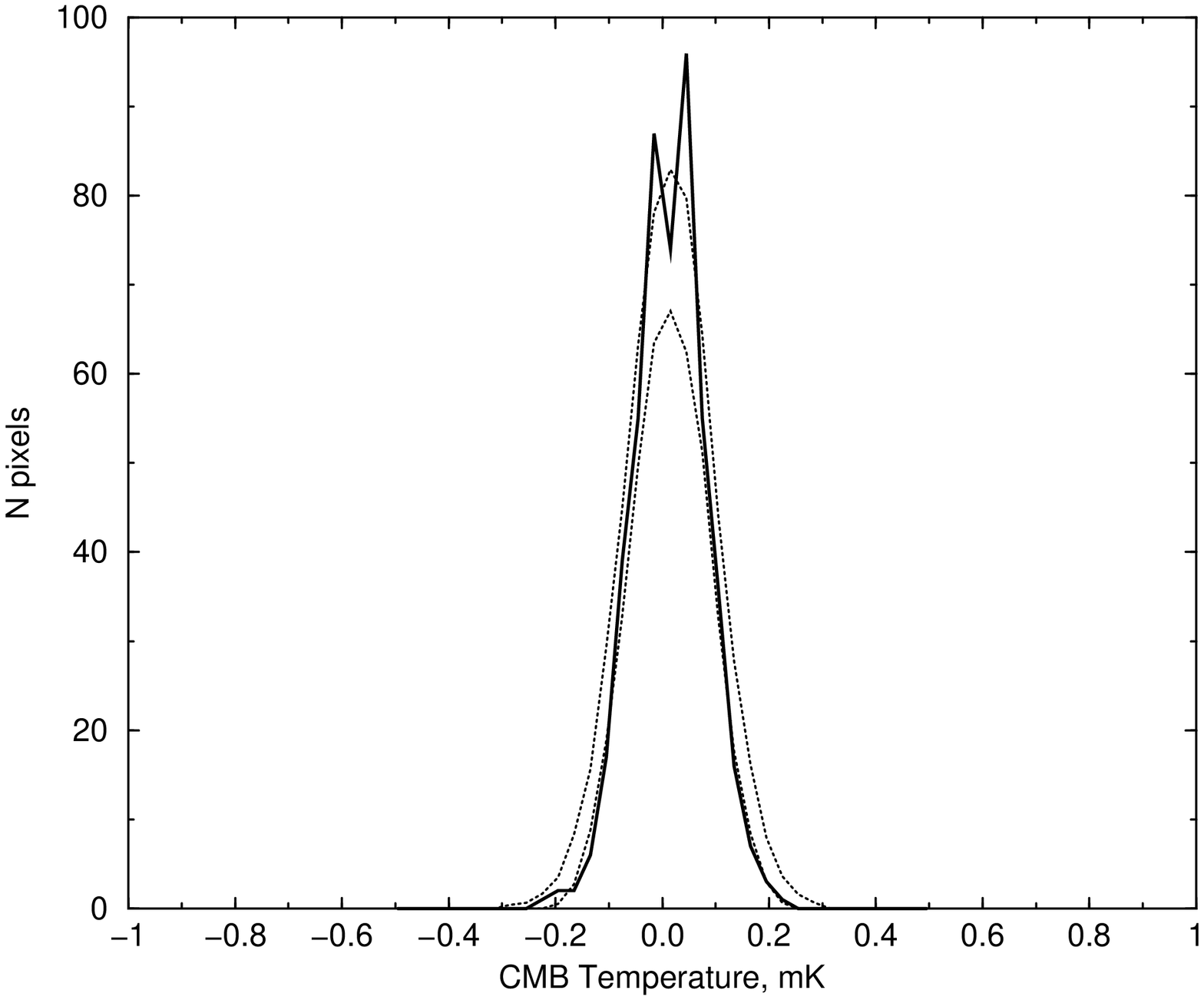,angle=-90,width=7cm}
\psfig{figure=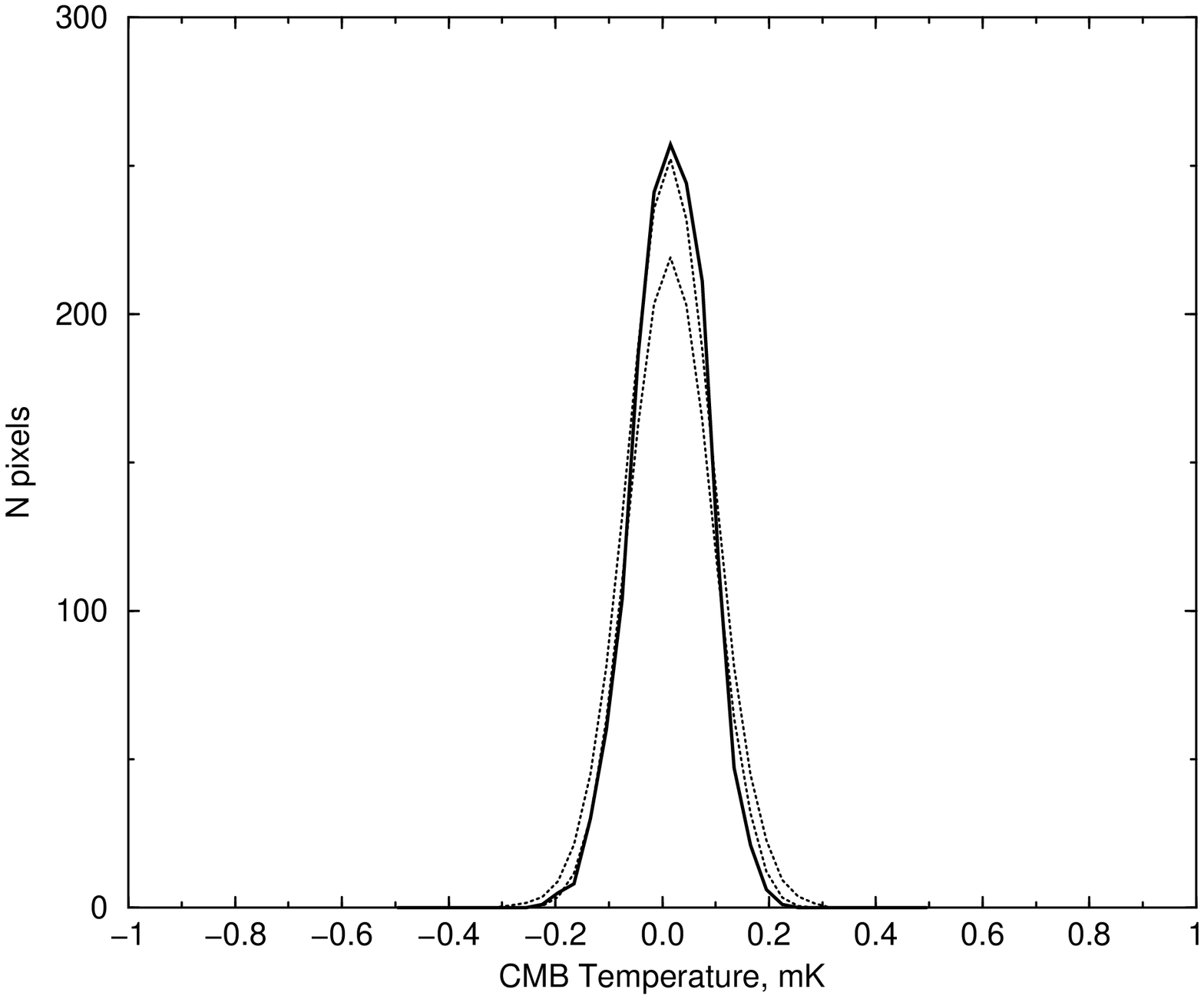,angle=-90,width=7cm}
}}}
\caption{ Distribution of the CMB fluctuations in the WMAP
pixels corresponding to GRB positions on the maps with a
resolution of $\ell_{max}=150$. The top left and top right panels
show the BeppoSAX data for the bursts with  $t<2$\,s and $t>2$\,s,
respectively. The bottom left and bottom right panels show the
data for the BATSE bursts  with $t<2$\,s and  $t>2$\,s,
respectively. The dashed lines show the 1$\sigma$ scatter of CMB
values in the $\Lambda$CDM cosmological model.}
\label{f6}
\end{figure*}

Figures~5 and 6 show the distributions of the CMB fluctuations for
four gamma-ray burst subsamples and two CMB maps with different
resolutions. The dashed lines show the expected 1$\sigma$ scatter
of CMB values in the  $\Lambda$CDM cosmological model. The bottom
left panel in Fig.~6, which shows the distribution of fluctuations
on the CMB map with a resolution of $\ell_{max}=150$ in the areas
of short gamma-ray bursts of the  BATSE catalog, stands out
conspicuously. In the positive part of the plot there is a peak,
which makes the distribution deviate from Gaussianity. Figure~7
shows the positions of these bursts superimposed onto the CMB
quadrupole. The statistical significance of this feature, which we
estimated by generating 10000 simulated CMB realizations with the
power spectrum corresponding to the $\Lambda$CDM model, is equal
to \mbox{$7\times10^{-4}$.}

\begin{figure*}
\centerline{
\psfig{figure=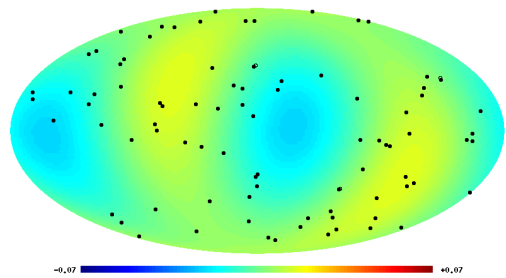,angle=-90,width=7cm,%
bblly=0pt,bbllx=0pt,bburx=500pt,bbury=840pt,clip=}
}
\caption{Positions of short BATSE bursts corresponding to the excess in the histogram
in Fig.\,6 (the bottom left panel) on the map of the quadrupole ILC component.}
\label{f7}
\end{figure*}

To analyze the distribution of GRB positions corresponding to the
excess peak, we used the GLESP software package \cite{glesp2} to
pixelize the map of burst positions. We chose the pixel size
700\arcmin$\times$700\arcmin\ in order to make the maximum pixel
value (the number of events inside the corresponding area) no less
than three, and to provide a significant dynamic range for the
harmonic analysis. Figure~8 shows the map of selected bursts
pixelized in such a way. The distribution of events on the sphere
shown in the image is clearly non-uniform: it is concentrated near
the ecliptic and/or equatorial poles, which is immediately
apparent in the smoothed map (Fig.\,9). Figure~9 also shows the
ecliptic and equatorial coordinate grids, and it is evident from
this figure that the asymmetry shows up both in the distribution
of the signal power with respect to the equatorial plane, and in
the number of spots that concentrate in the Southern Hemisphere in
both coordinate systems. Note also that the hot spot above the
Galactic center lies in the equatorial plane.

\begin{figure*}
\centerline{
\psfig{figure=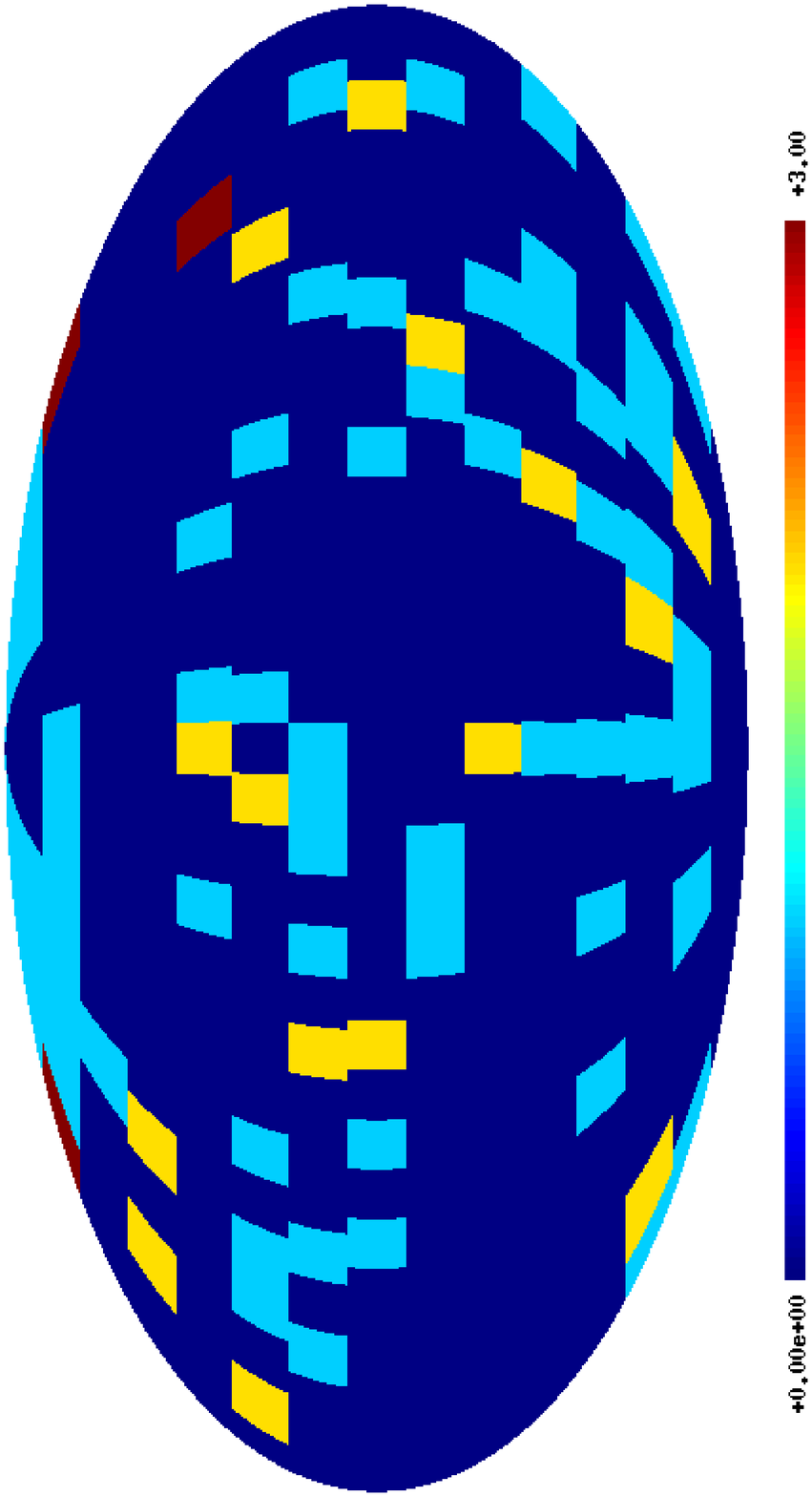,angle=-90,width=7cm}
}
\caption{Pixelized map of positions of short gamma-ray bursts of
the BATSE catalog corresponding to the excess in the histogram in
Fig.\,6 (the bottom left panel). The pixel size
700\arcmin$\times$700\arcmin\ is chosen in a way to make the
maximum pixel value---the number of events inside the
corresponding sky area---greater than or equal to three.}
\label{f8}
\end{figure*}

\begin{figure*}
\centerline{\hbox{
\psfig{figure=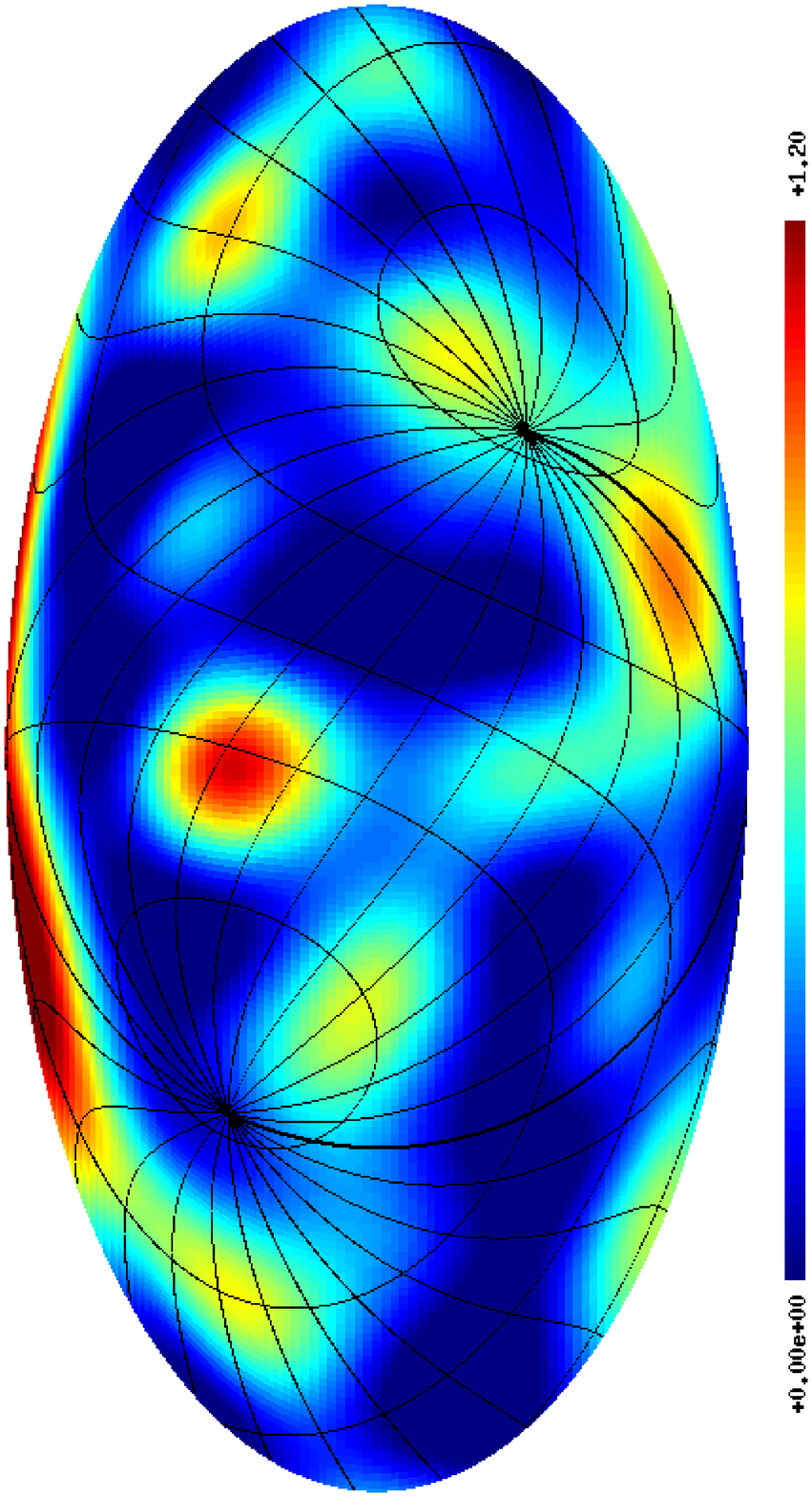,angle=-90,width=7cm}
\psfig{figure=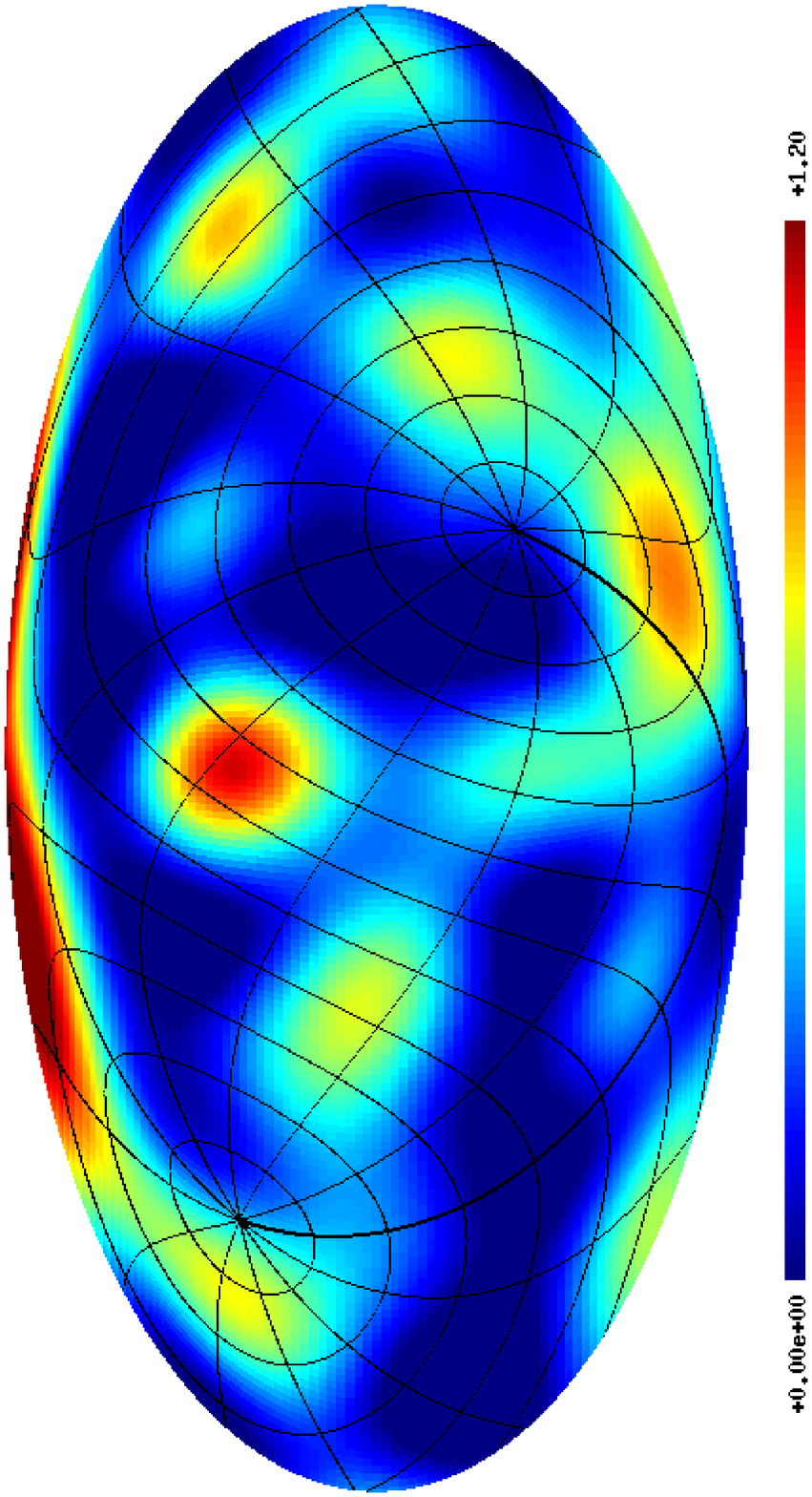,angle=-90,width=7cm}
}}
 \caption{Smoothed sky map corresponding to the pixelization used in Fig.\,8,
with the ecliptic (right) and equatorial (left) coordinate grids superimposed.	}
\label{f9}
\end{figure*}

\begin{figure*}
\centerline{\hbox{
\psfig{figure=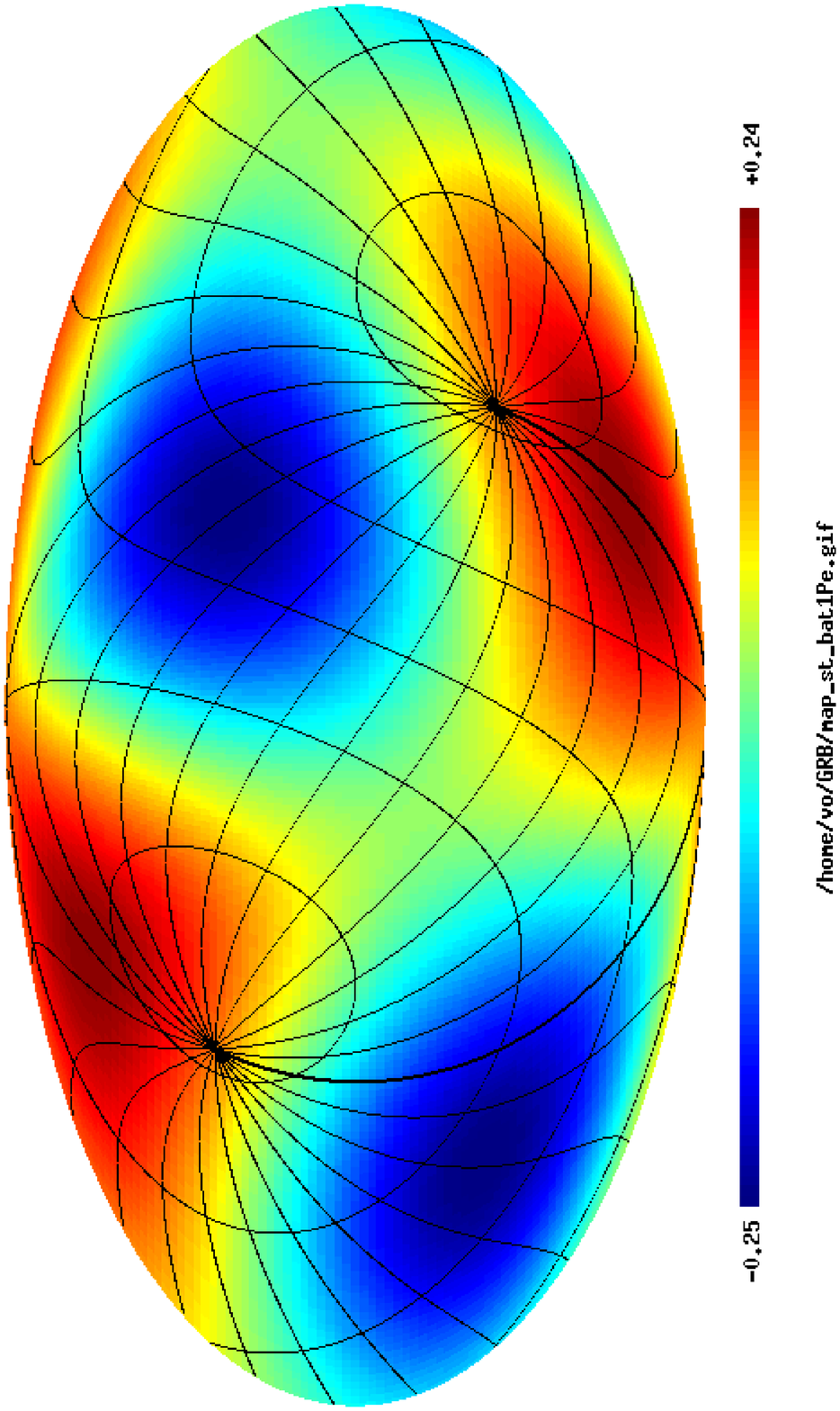,angle=-90,width=7cm,%
bblly=0pt,bbllx=0pt,bburx=500pt,bbury=840pt,clip=}
\psfig{figure=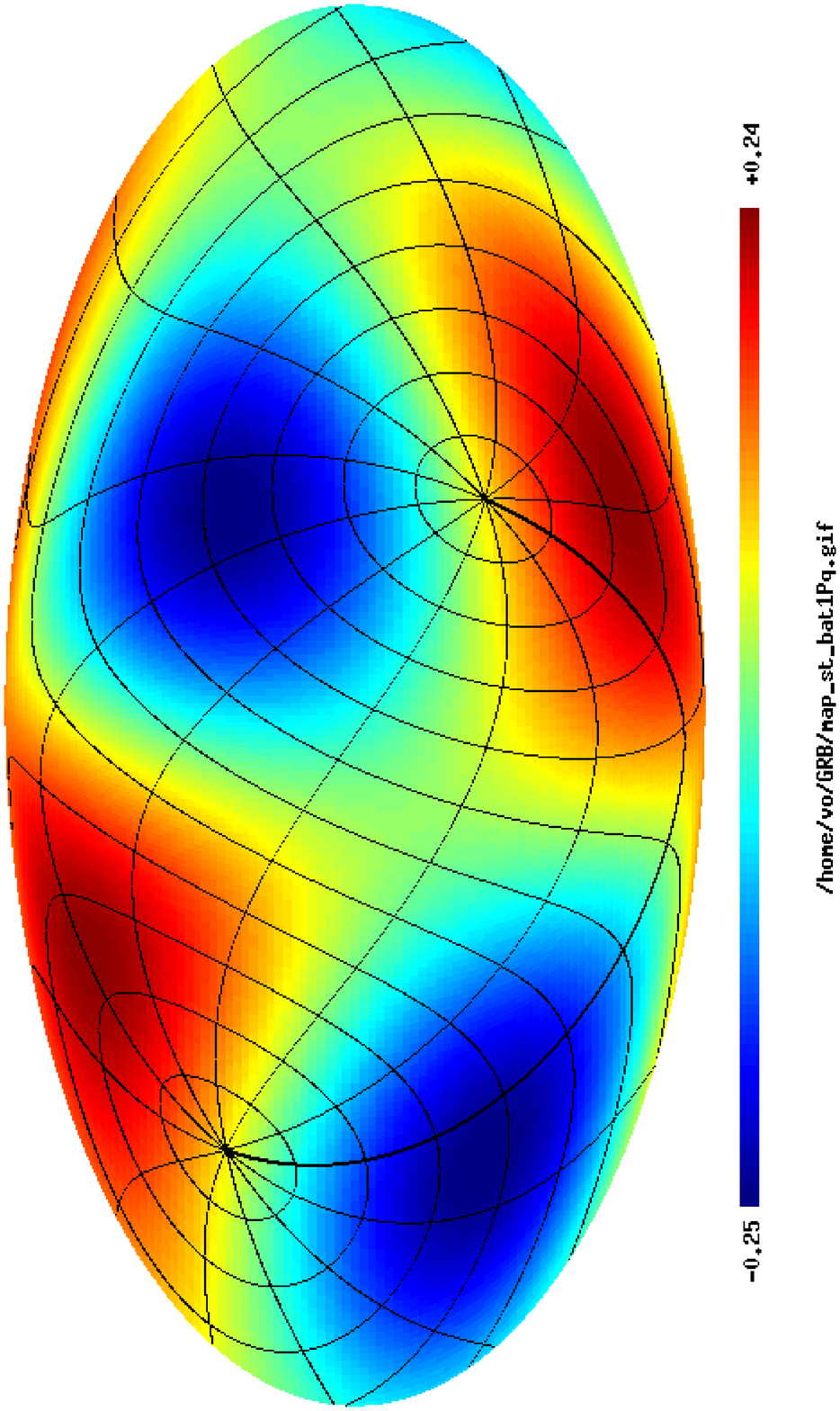,angle=-90,width=7cm,%
bblly=0pt,bbllx=0pt,bburx=500pt,bbury=840pt,clip=}
}}
 \caption{A quadrupole of the smoothed map  (Fig.\,9)
with the ecliptic (right) and equatorial (left) coordinate grids superimposed. }
\label{f10}
\end{figure*}

To reveal the asymmetries more clearly, we demonstrate a
quadrupole of the smoothed map (Fig.\,10) with the ecliptic and
equatorial coordinate grids superimposed, as we do it in Fig.\,9.
It is evident from the positions of quadrupole spots that: (1) the
poles of both coordinate systems are located on the outskirts of
hot spots; (2) cold spots are located symmetrically with respect
to the equatorial planes, and (3) the Galactic center is located
in the saddle-shaped area between cold and hot spots.

Our analysis of gamma-ray burst positions in the area of the peak
in the distribution of short BATSE bursts with respect to the CMB
revealed their unexpected sensitivity to local (near-Earth)
coordinate systems. To analyze this problem in more detail, we
pixelize all the four subsamples of the GRB catalogs using the
method of mosaic correlation described in the previous section.

\section{GRB AND WMAP ILC DISTRIBUTION CORRELATION MAPS}

To verify and refine the correlation properties of the maps of GRB
positions and CMB fluctuations, we pixelized the maps of GRB
positions for four gamma-ray burst subsamples (Fig.\,11). Like at
the previous stage, we chose the pixel size
200\arcmin$\times$200\arcmin\, ($\ell_{max}=26$) in order to
ensure that the maximum pixel value---the number of events in the
corresponding area---would be not less than three.

\begin{figure*}
\centerline{\vbox{
\hbox{
\psfig{figure=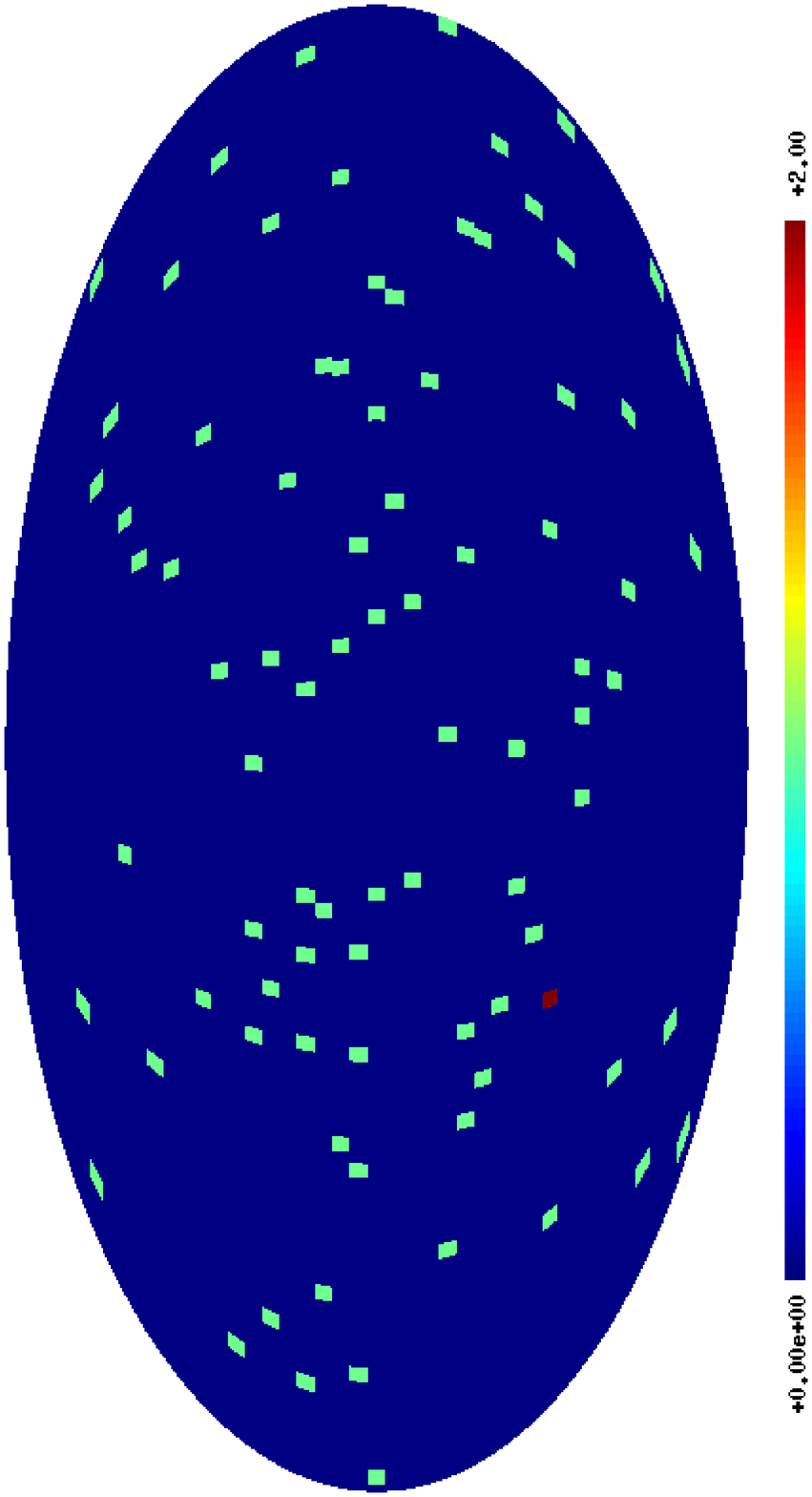,angle=-90,width=7cm}
\psfig{figure=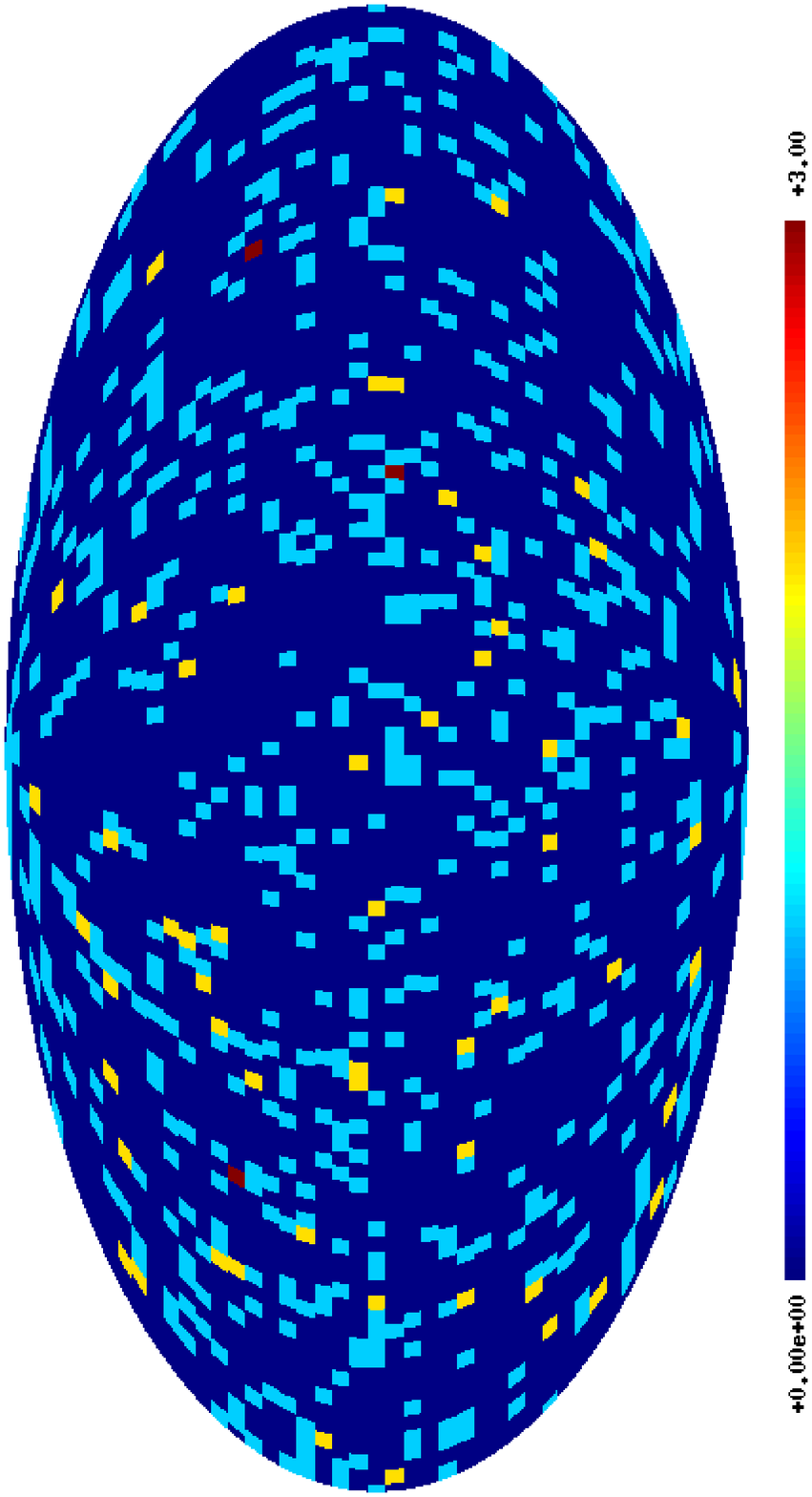,angle=-90,width=7cm}
} \hbox{
\psfig{figure=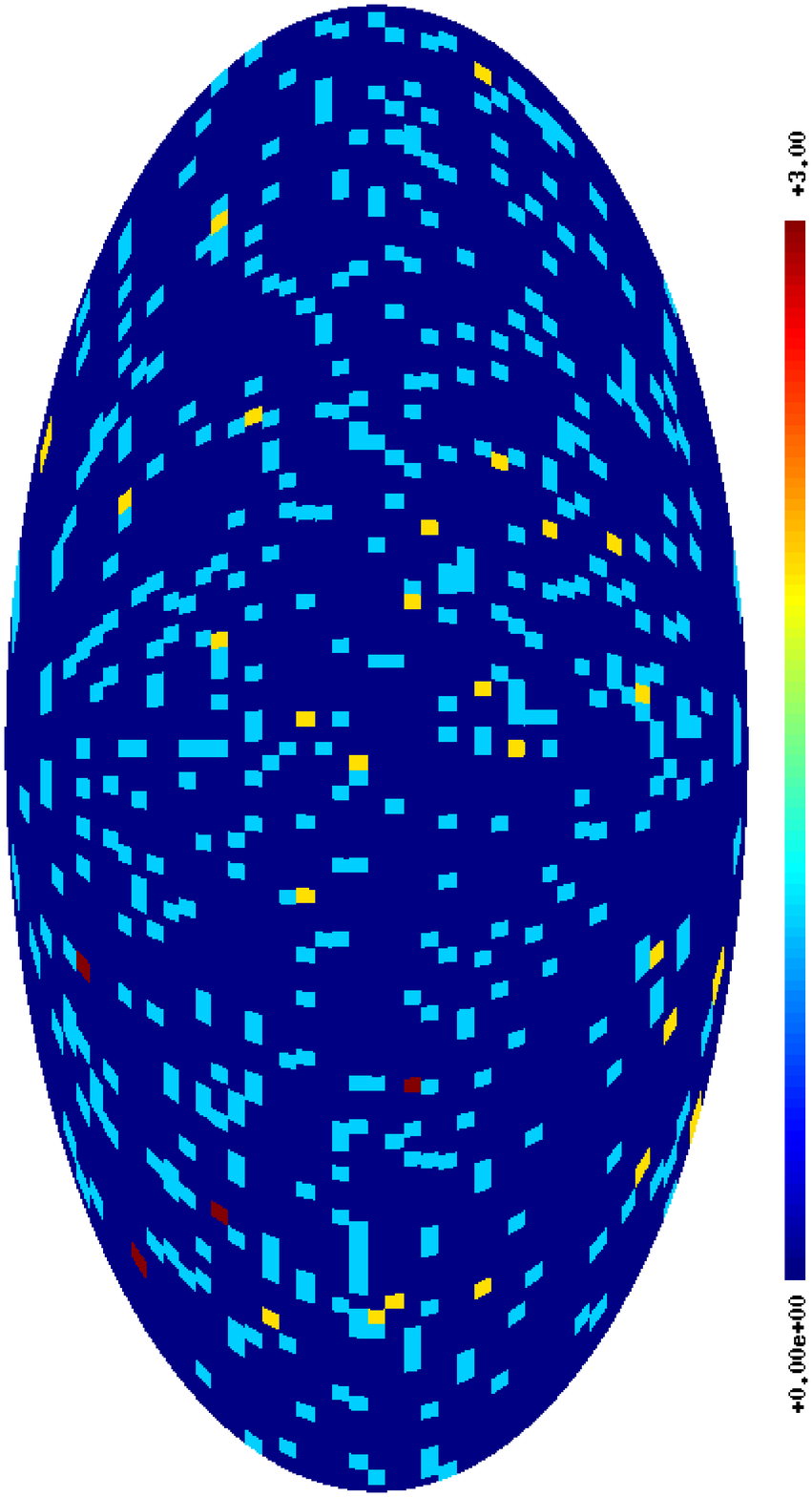,angle=-90,width=7cm}
\psfig{figure=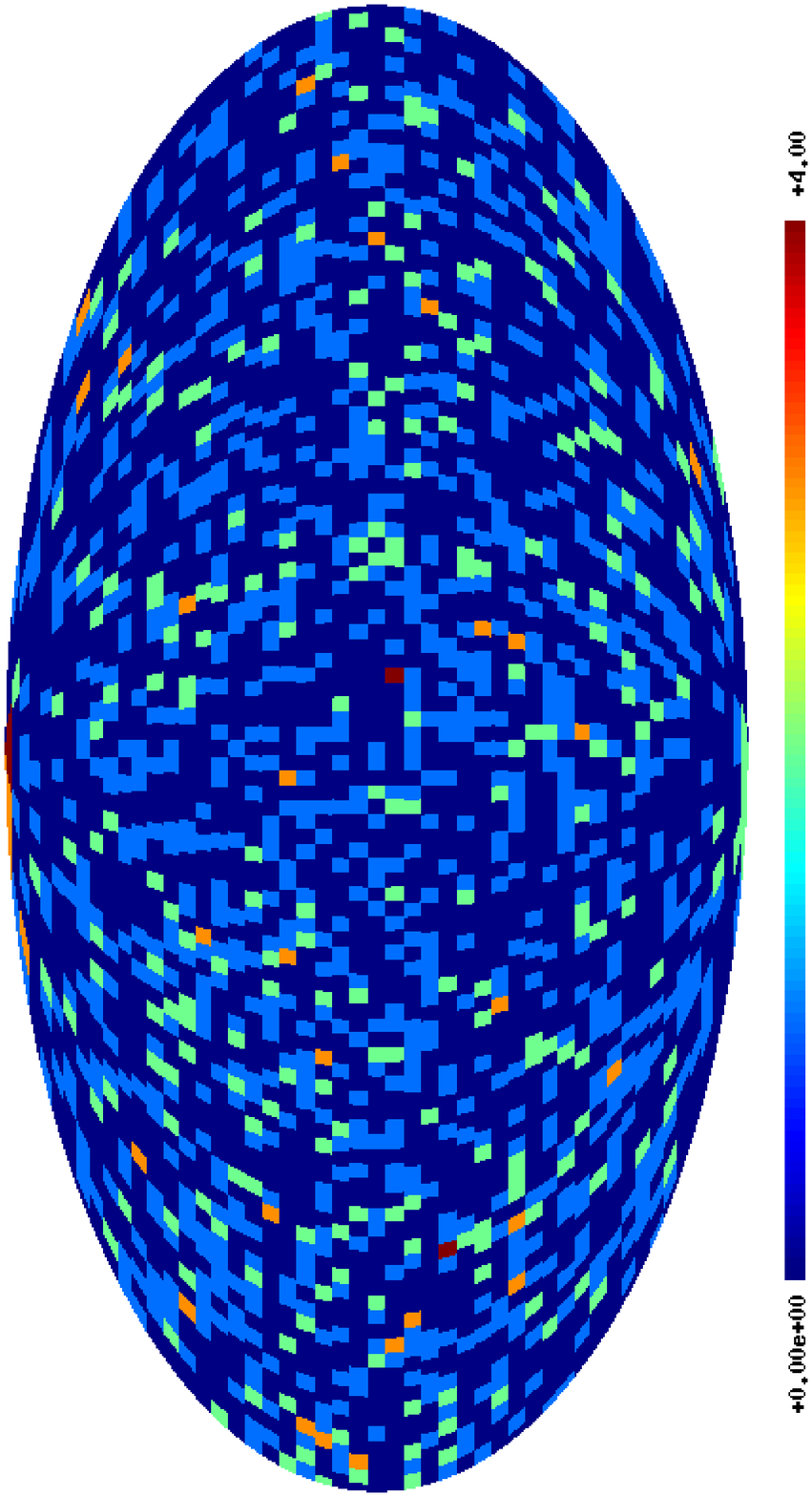,angle=-90,width=7cm}
}}}
 \caption{Pixelized maps of event positions in gamma-ray burst subsamples. The pixel
size is 200$'\times200'$.  The top left and top right panels show
the BeppoSAX data for bursts with  $t<2$\,s and	 $t>2$\,s,
respectively. The bottom left and bottom right panels show the
data for the BATSE bursts  with	 $t<2$\,s and  $t>2$\,s,
respectively. }
\label{f11}
\end{figure*}

We then performed the mosaic correlation for the BATSE--BeppoSAX
and BATSE--CMB pairs (Fig.\,12). The pixel size for these
correlations was equal to 500\arcmin$\times$500\arcmin.

\begin{figure*}
\centerline{\vbox{
\hbox{
\psfig{figure=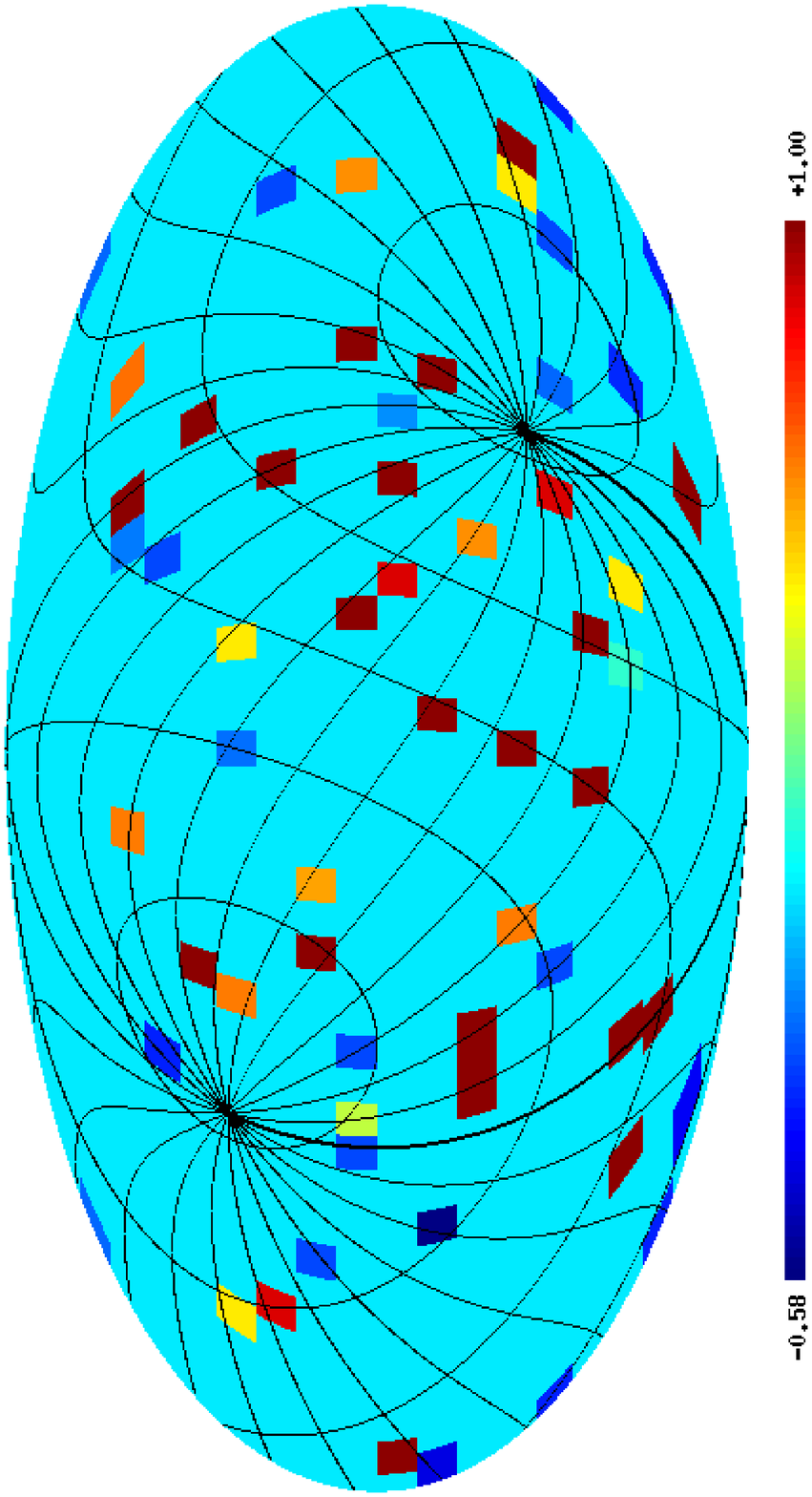,angle=-90,width=7cm}
\psfig{figure=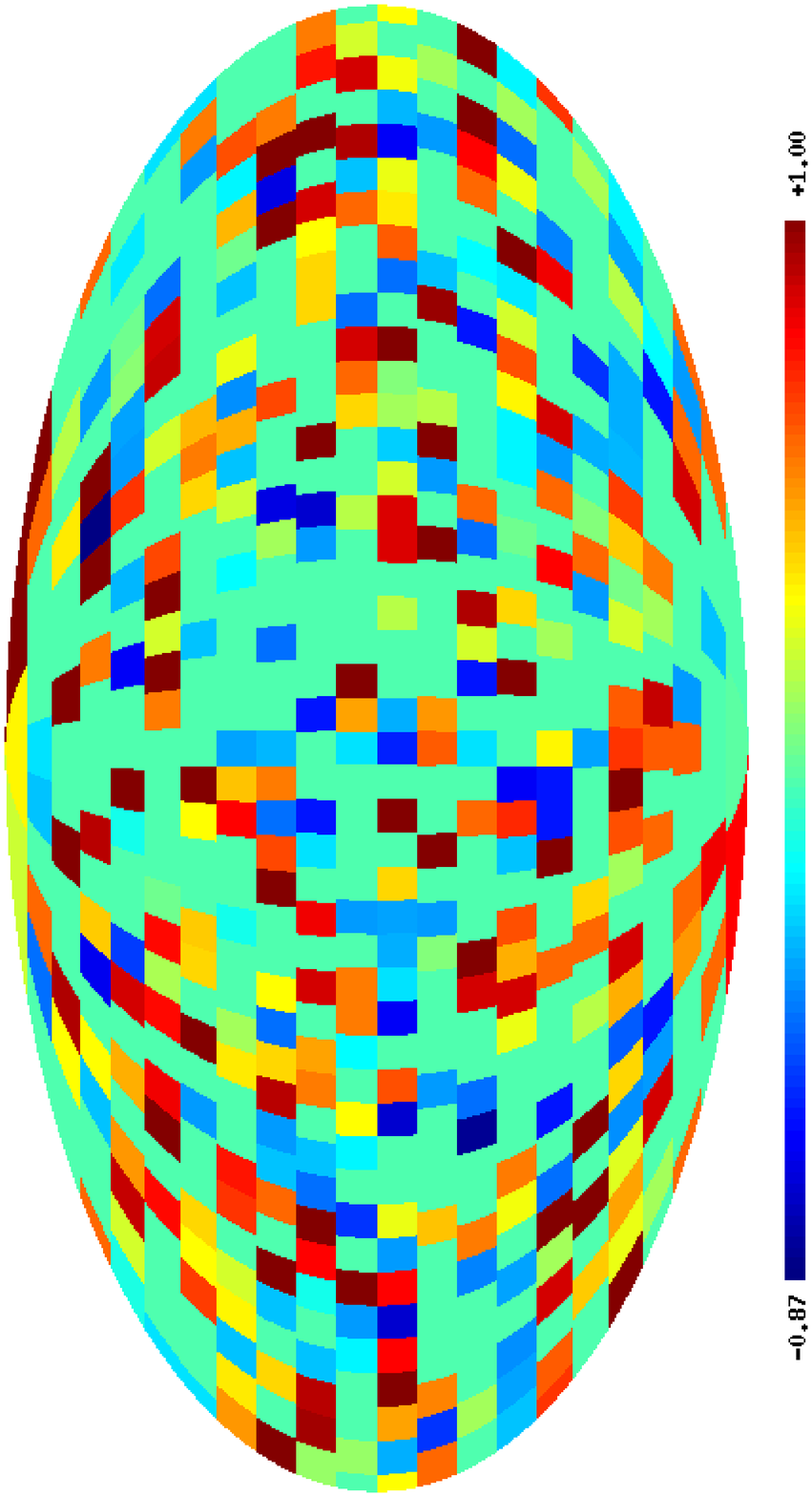,angle=-90,width=7cm}
} \hbox{
\psfig{figure=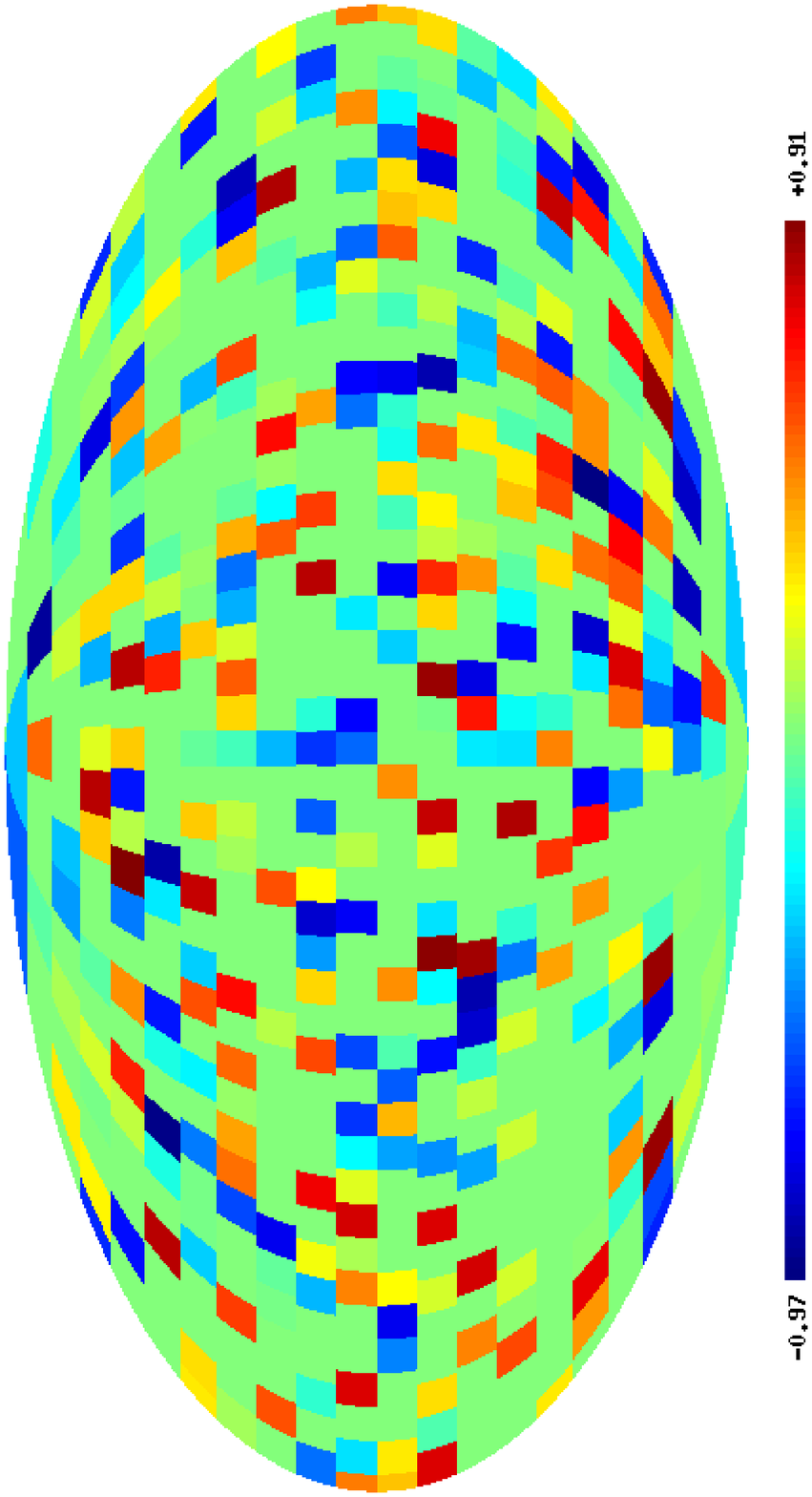,angle=-90,width=7cm}
\psfig{figure=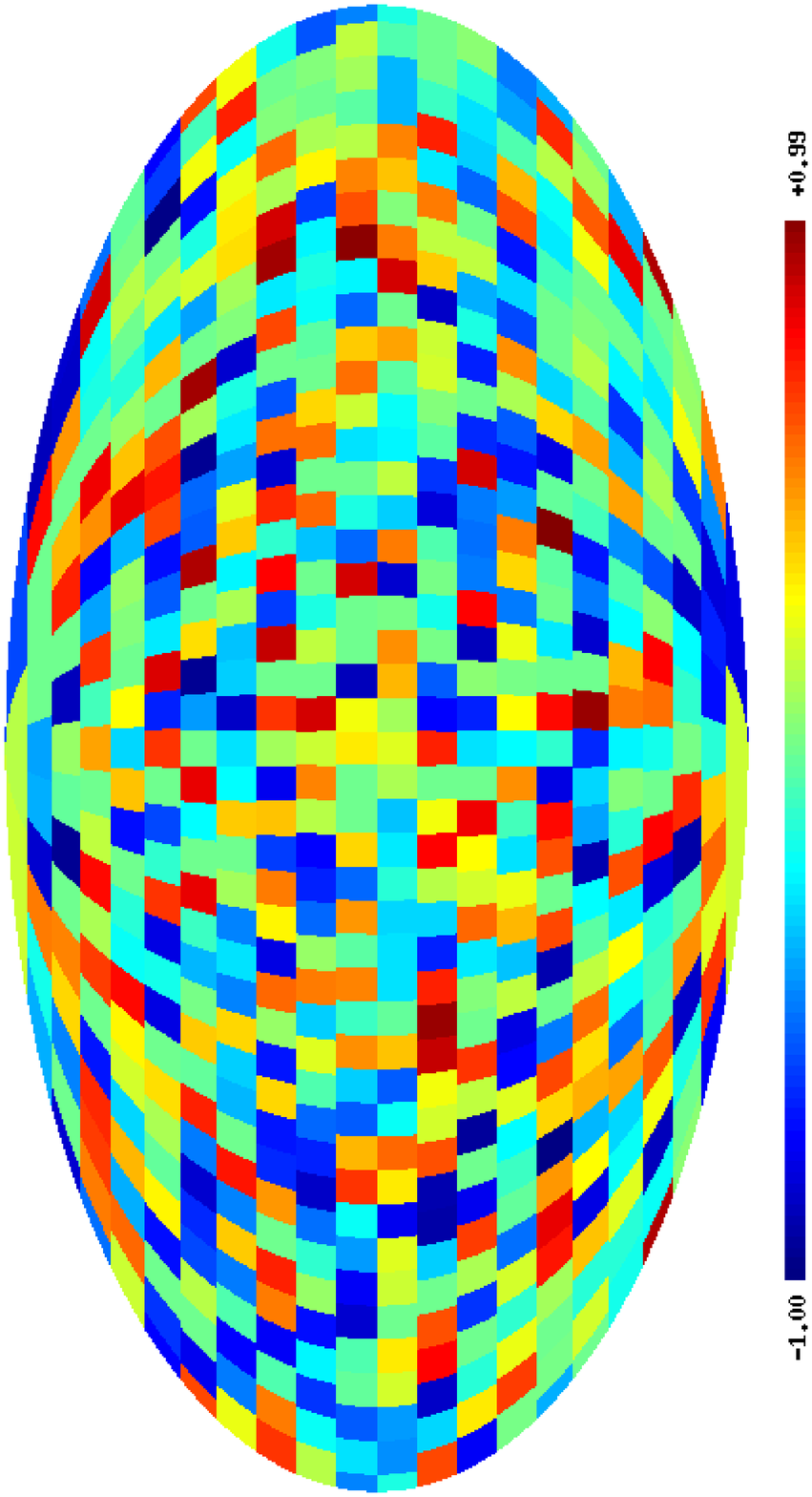,angle=-90,width=7cm}
}}}
 \caption{Correlation maps for the CMB data ($\ell_{max}=26$) and gamma-ray burst positions
in the Galactic coordinate system. The pixel size for these correlations is
equal to 500$'\times500'$. The top left panel shows the correlation between BATSE and
BeppoSAX data, $t<2$\,s, with the ecliptic coordinate grid superimposed.
The top right panel shows the correlation between  BATSE ($t<2$\,s) and CMB data.
The bottom left panel shows the correlation between BATSE and BeppoSAX data, $t>2$\,s.
The bottom right panel shows the correlation between  BATSE ($t>2$\,s) and CMB
data.  }
\label{f12}
\end{figure*}

We can now use the expansion defined by formula (\ref{eq1}) to
compute the angular power spectrum \mbox{of the map:}
\begin{equation}
C(\ell) =
   \frac{1}{2\ell+1}\left[|a_{\ell 0}|^2 +2\sum_{m=1}^\ell
     |a_{\ell,m}|^2\right]\,.
\end{equation}

The power spectrum allows us to identify the main harmonics
contributing to the correlation map. Figure~13 shows the power
spectra of the correlation coefficient maps (mosaic correlation)
between the BATSE and CMB data ($\ell_{max}=26$) and between the
BATSE and BeppoSAX gamma-ray positions.

\begin{figure*}
\centerline{\vbox{
\hbox{
\psfig{figure=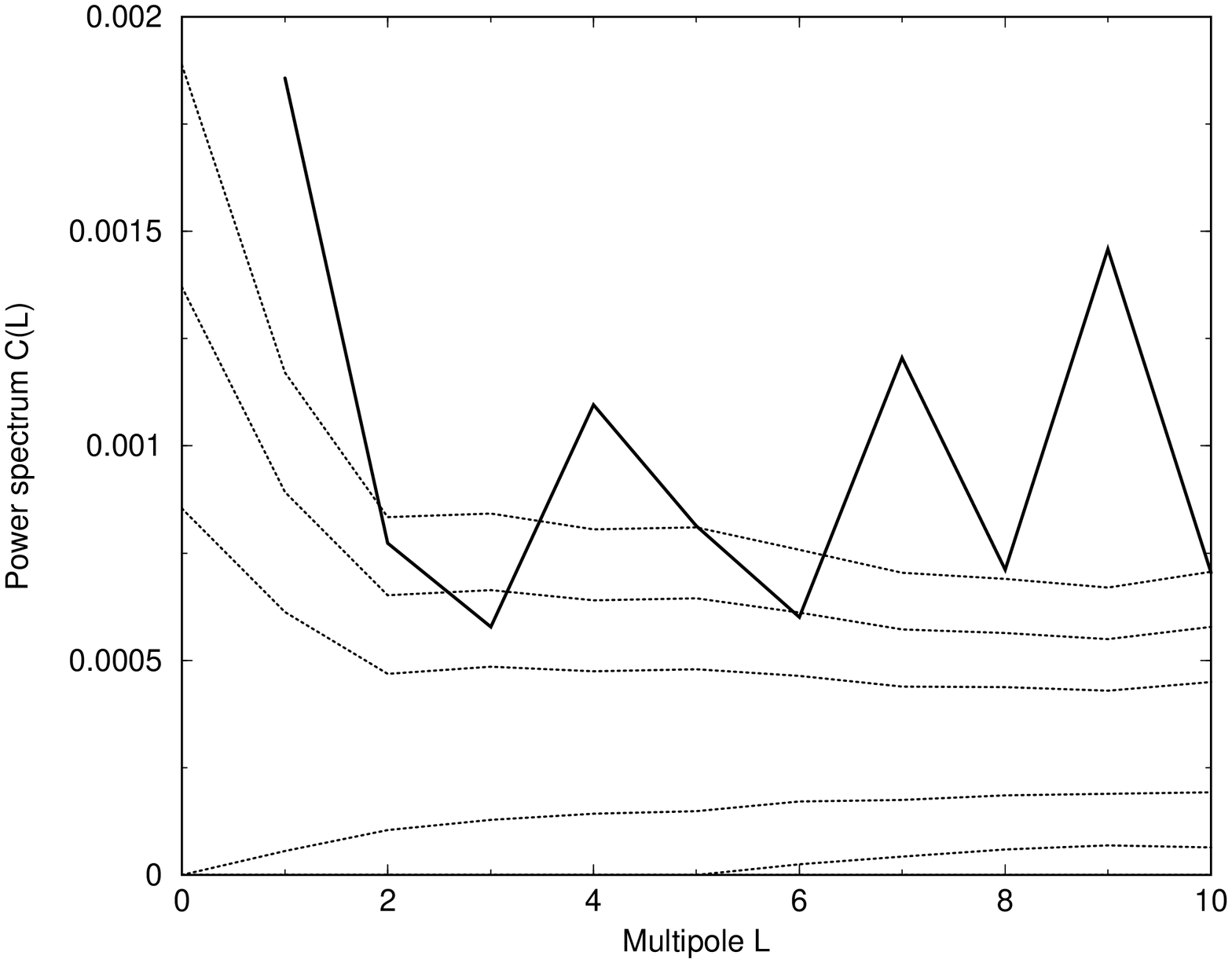,angle=-90,width=7cm}
\psfig{figure=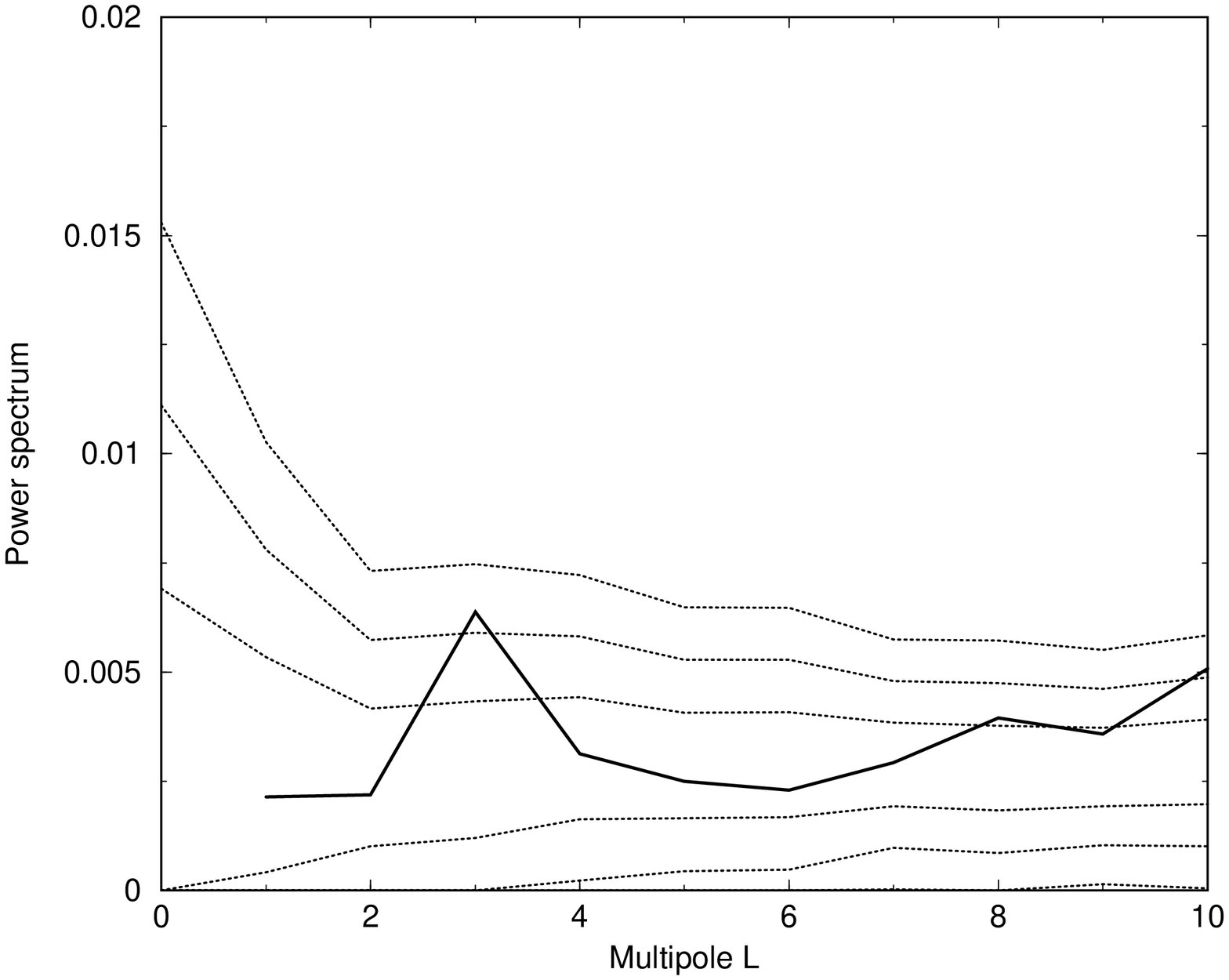,angle=-90,width=7cm}
} \hbox{
 \psfig{figure=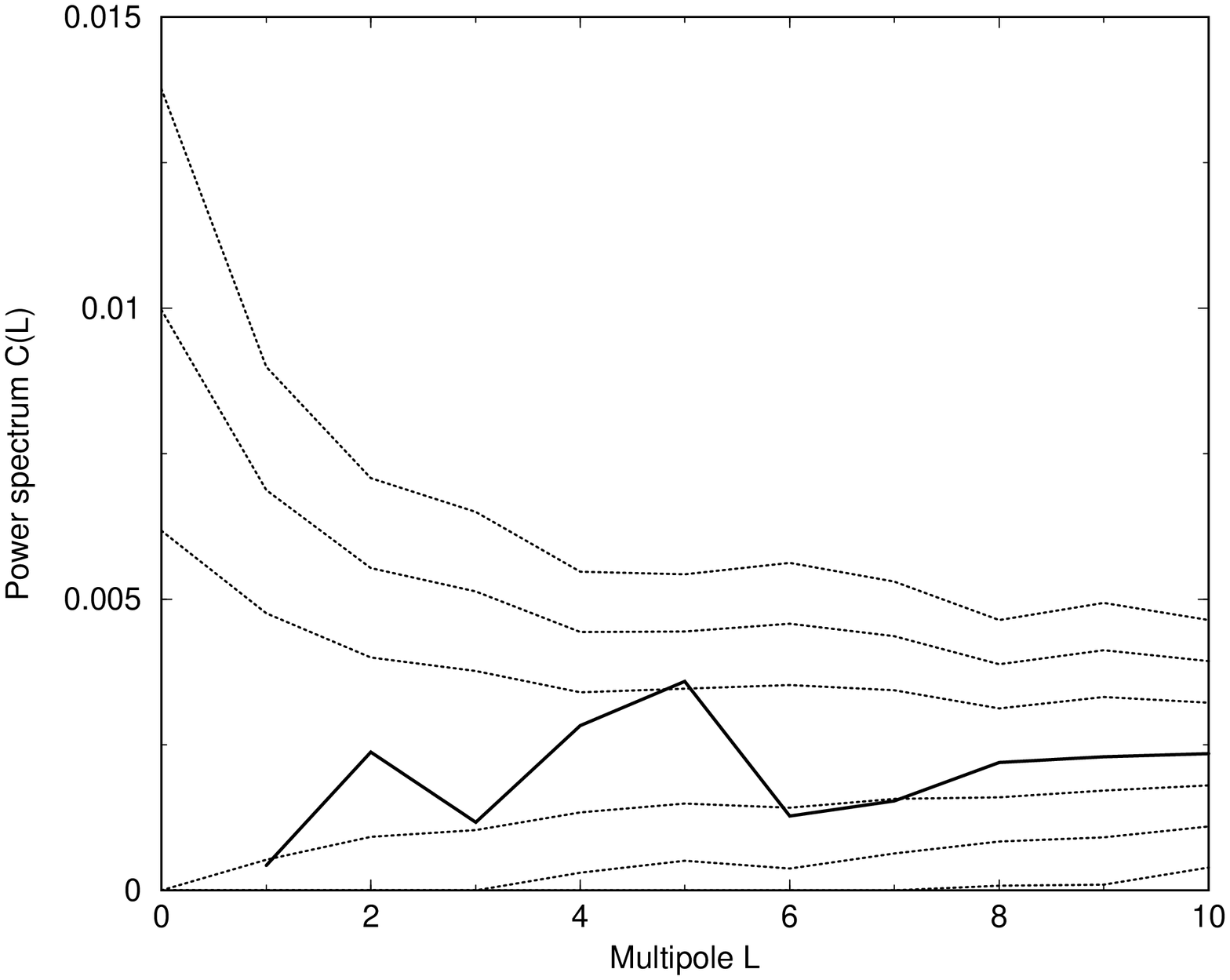,angle=-90,width=7cm}
\psfig{figure=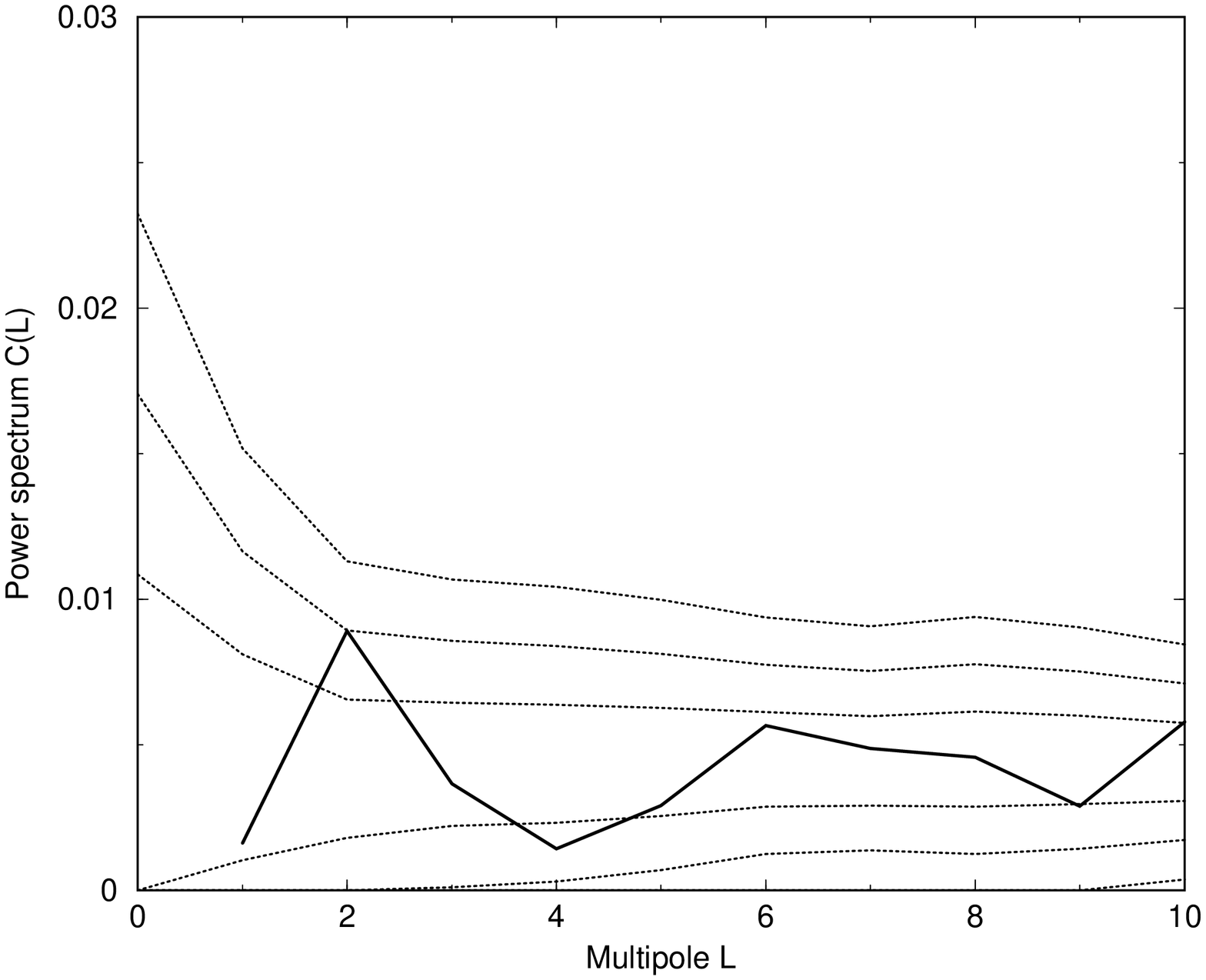,angle=-90,width=7cm}
} }} \caption{Power spectra of the correlation coefficient maps
between the BATSE GRB positions and BeppoSAX and CMB data
($\ell_{max}=26$) (the solid line). The pixel size of the
correlations is 500\arcmin$\times$500\arcmin. The top left panel
shows the spectrum of correlation between the  BATSE and BeppoSAX
data, $t<2$\,s. The top right panel shows the spectrum of
correlation between the BATSE ($t<2$\,s) and CMB data. The bottom
left panel shows the spectrum of correlation between the BATSE and
BeppoSAX data, $t>2$\,s. The bottom right panel shows the spectrum
of correlation between the BATSE ($t>2$\,s) and CMB data. The
dashed lines show the $\pm \sigma$, $\pm 2\sigma$, and $\pm
3\sigma$ deviations from the mean, where $\sigma$ is estimated
from the spectra of 200 simulated random realizations of the zero
hypothesis, which assumes a uniform distribution of gamma-ray
bursts in the sky and the Gaussian CMB amplitude distribution
corresponding to the $\Lambda$CDM cosmology. }
\label{f13}
\end{figure*}

\begin{figure}
\centerline{
\psfig{figure=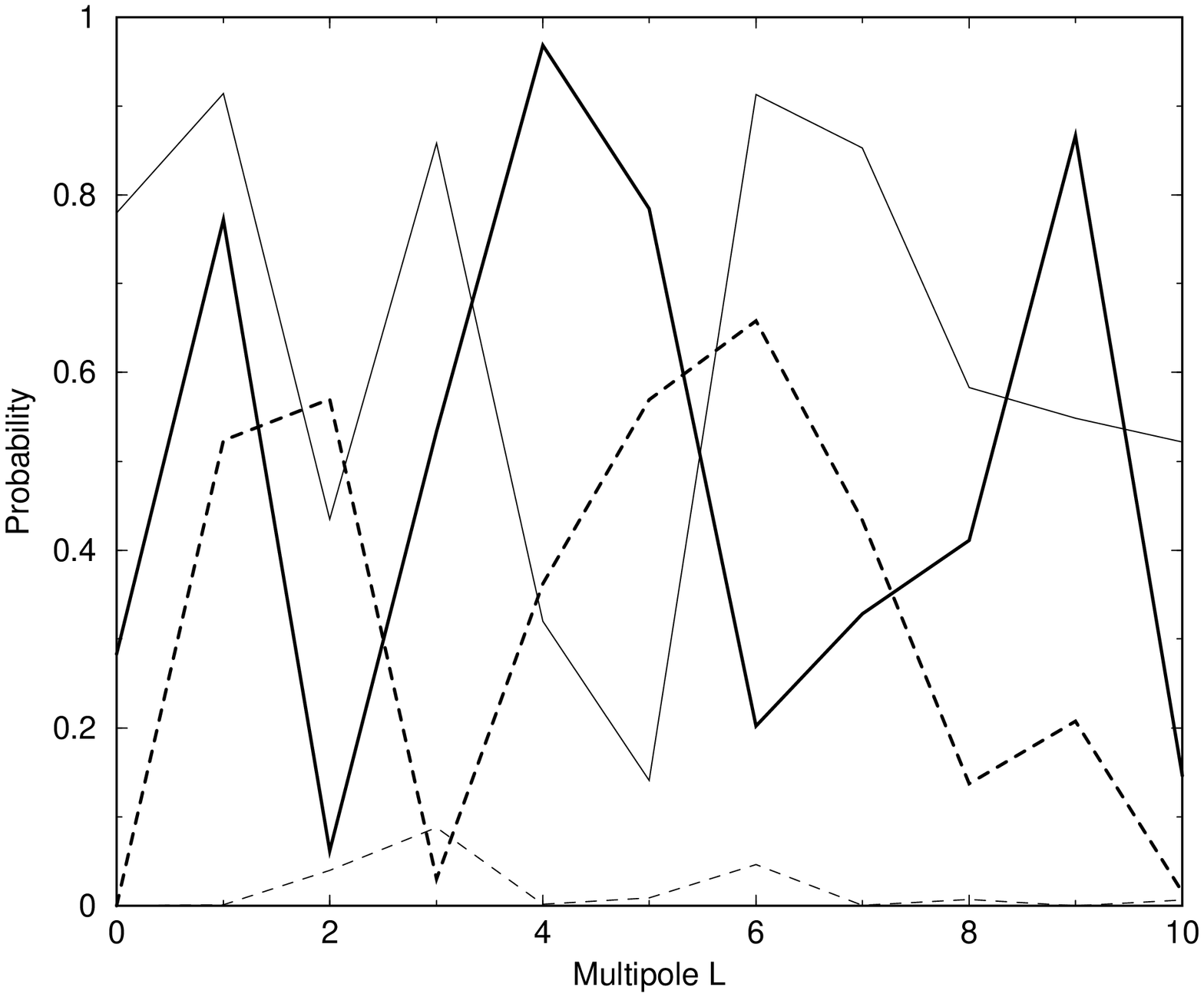,angle=-90,width=7cm}
}
 \caption{Significance levels for the deviations from the mean level for the power
spectra of the correlation maps: short gamma-ray burst
BATSE-BeppoSAX correlations (the thin dashed line); long gamma-ray
burst BATSE-BeppoSAX correlations (the thick dashed line); short
BATSE bursts and CMB fluctuations (the thin solid line), and long
BATSE bursts and CMB fluctuations (the thick solid line). The
statistical significance was estimated via direct numerical
simulations involving 10000 random realizations of the zero
hypothesis, which consists in the uniform GRB distribution in the
sky and the Gaussian CMB amplitude distribution corresponding to
the $\Lambda$CDM cosmology.}
\label{f14}
\end{figure}

Three correlation power spectra stand out among other
correlations: those between the positions of short BATSE and
BeppoSAX bursts (the top left panel in Fig.\,13) and between short
and long BATSE bursts and CMB fluctuations (top and bottom right
panels, respectively in Fig.\,13). The peaks of these spectra
exhibit features that formally lie beyond the  2$\sigma$ level,
and, in the case of the harmonics $\ell =1$, 4, 7, 9, even beyond
3$\sigma$. To estimate the statistical significance of these
deviations, we performed numerical simulations of the zero
\mbox{hypothesis} by generating: (1) 10,000 random realizations of
the set of sky positions of the BeppoSAX gamma-ray bursts assuming
that they are uniformly distributed in the sky and (2) random CMB
maps with the Gaussian amplitude distribution corresponding to the
$\Lambda$CDM cosmology, and computed the correlations and the
corresponding power spectra for these simulated data in the same
way as we did it for real data. Figure~14 shows the probabilities
for a simulated realization to reach the deviation levels of the
features observed in Fig.\,13 within the framework of the zero
\mbox{hypothesis} i.e., the significance levels of the peaks in
the power spectrum. The resulting significance levels for the
multipoles  $\ell =1$, 4, 7, and 9 of the correlation maps between
the short BATSE and BeppoSAX bursts are equal to 0.0011, 0.0012,
0.0001, and less than 0.0001, respectively. At the same time, the
significance level for the octupole of the correlations between
the long  BATSE and BeppoSAX bursts is equal to 0.0325, and that
for the quadrupole of the correlation map between the long  BATSE
bursts and CMB fluctuations is equal to 0.0614. Figure~15 shows
the map of one of the harmonics ($\ell=7$) in the distribution of
the correlation coefficients between the positions of short BATSE
and BeppoSAX bursts. Figure~16 shows the maps of the peak
harmonics ($\ell=3$ for short and $\ell=2$ for long bursts) in the
spectrum of the maps of correlations between the BATSE and CMB.

\begin{figure}
\centerline{
\psfig{figure=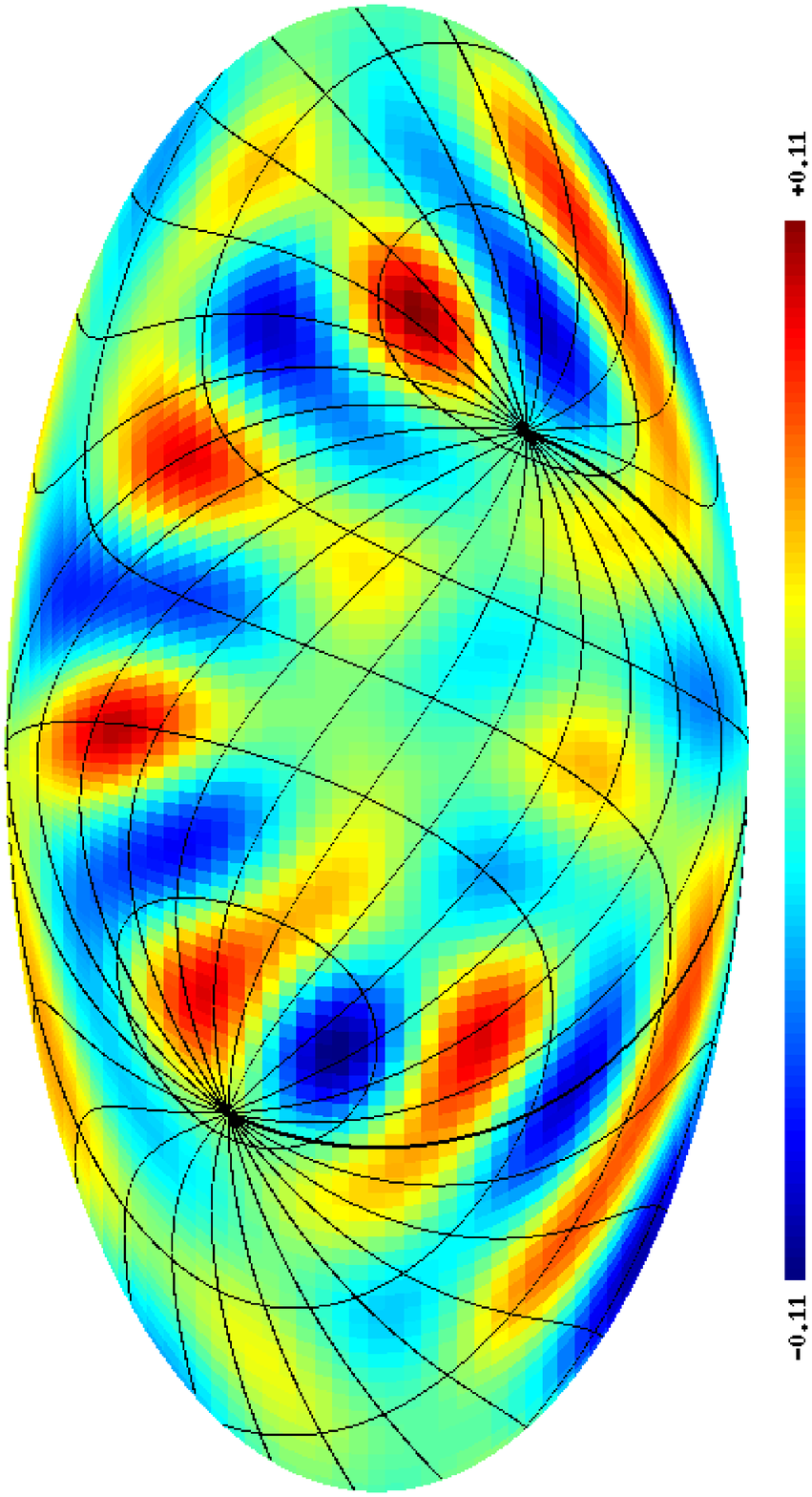,angle=-90,width=7cm}
}
 \caption{Map of the seventh harmonic ($\ell=7$) in the expansion of the map of correlations
between the positions of short BATSE and BeppoSAX bursts with the equatorial coordinate
grid superimposed.  }
\label{f15}
\end{figure}

\begin{figure*}
\centerline{\vbox{
\psfig{figure=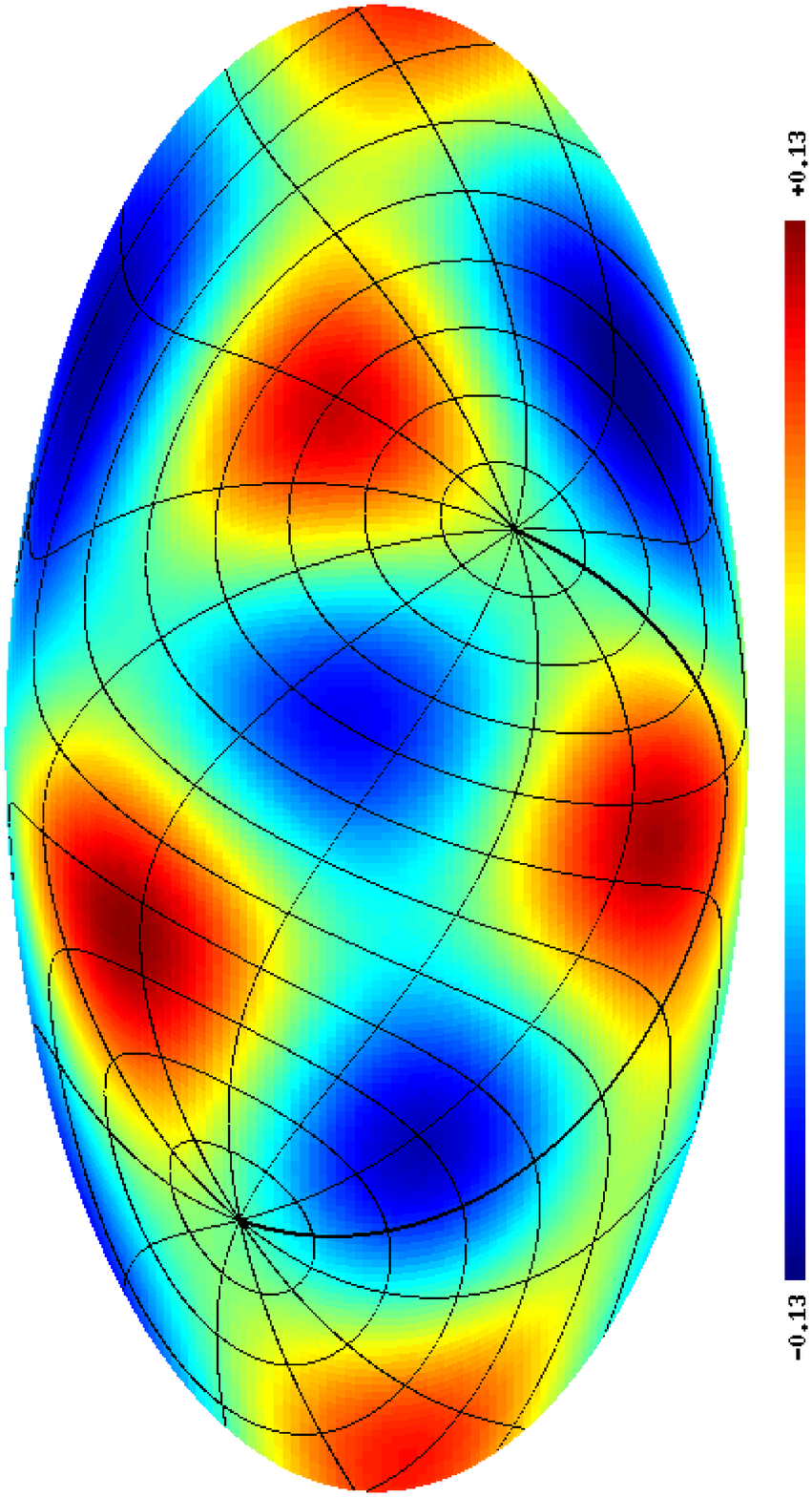,angle=-90,width=7cm}
\psfig{figure=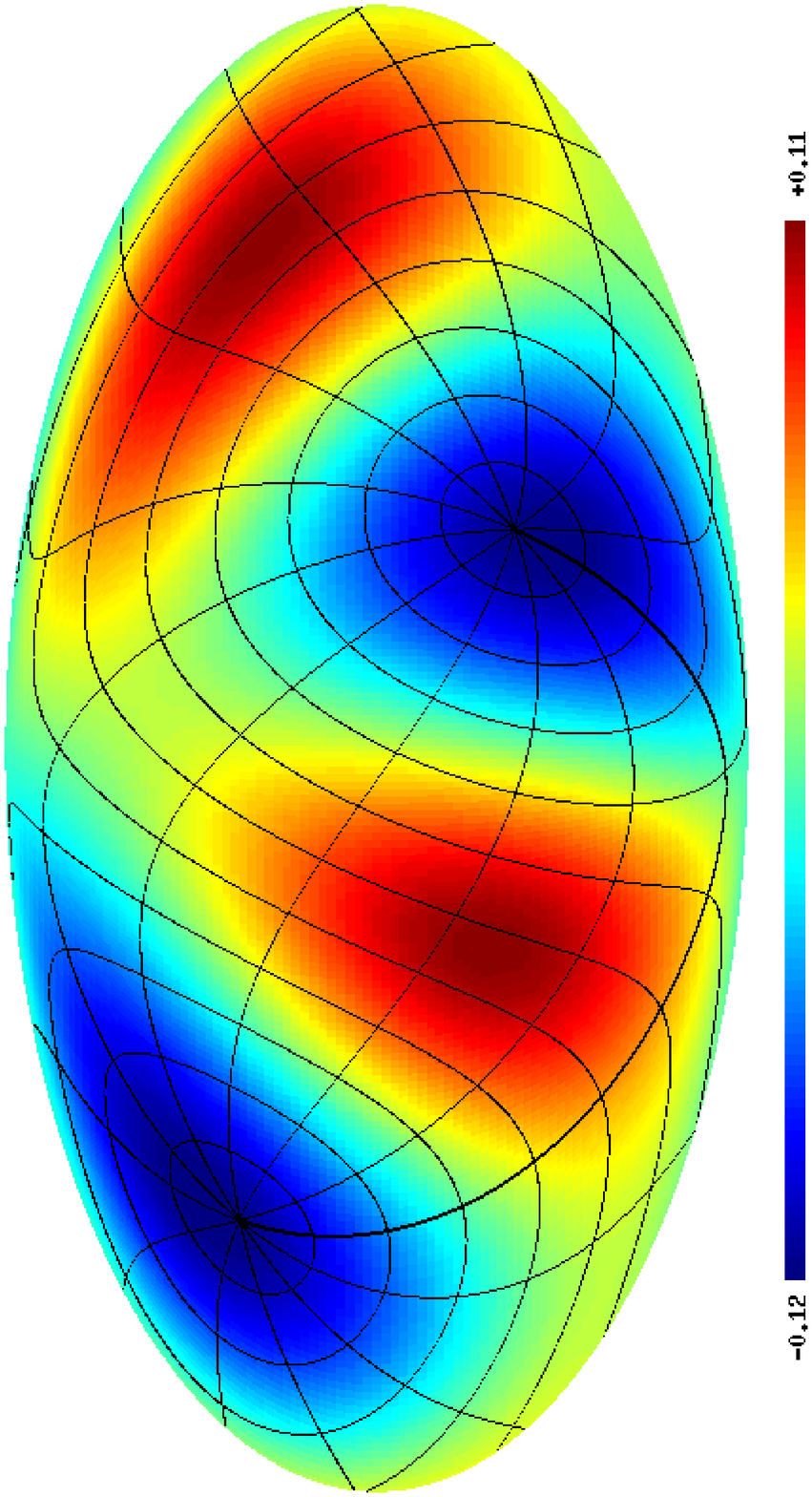,angle=-90,width=7cm}
}}
 \caption{Maps of the harmonics identified in the power spectrum (Fig.\,13). Correlation maps
between the BATSE burst positions ($t>2$ s) and the BeppoSAX data
(the top panel, $\ell=3$) and between the BATSE burst positions
($t>2$ s) and the CMB (the bottom panel, $\ell=2$) with the
equatorial coordinate grid superimposed.  }
\label{f16}
\end{figure*}

A distinguishing feature of the maps shown with superimposed
coordinate grids  are the positions of the poles of coordinate
systems. It is evident from Fig.\,15 that the ecliptic poles are
located at special \mbox{points---in} the saddles between the
maxima and minima of the signal distribution on the map.
Similarly, the equatorial poles are located in the saddles in the
map of the correlation octupole (the left panel in Fig.\,16). We
can see in the right panel (the quadrupole) of Fig.~16 that the
poles are located in the minima of the quadrupole. We use the
method described above based on the generation of 10,000 random
realizations of the GLESP pixelization with 102 pixels at the
equator to estimate the statistical significance of such a
configuration i.e., the probability for the minima of the
quadrupole to occur in the 5-degree radius areas centered on the
equatorial poles, and find it to be equal to 0.0035.

\section{DISCUSSION}

Our analysis of correlations between the maps of GRB positions and
the CMB shows that such correlations do exist. Of great interest
is the orientation (phase properties) of correlation maps, where
we can point out two points: (1)  in the case of the correlations
between the short GRB positions in different catalogs and their
correlations with the CMB, the features found (the positions of
the poles) are observed both in the equatorial and ecliptic
coordinate systems; moreover, despite a small number of short
BeppoSAX and BATSE events, the correlations between their
positions in 500\arcmin$\times$500\arcmin\ windows exhibit a chain
of events in the ecliptic plane (the top left panel in Fig.\,12),
and a predominant occurrence of the correlated pixels in the
Southern Hemisphere; (2) the correlations between the long BATSE
events and CMB fluctuations exhibit features in the equatorial
coordinate system, while the joint probability of high quadrupole
amplitude and of the quadrupole minimum occurrence at the
equatorial pole is practically equal to zero in the case of
correlation with random maps.

The correlations found between the GRB positions and the CMB, that
are sensitive to the equatorial coordinate system must be due to
the systematic effects. This way, a relatively greater number of
gamma-ray events near the equatorial poles may be due to
longer exposures of the satellite cameras due to the observing
method employed. Both satellites that observed the gamma-ray
bursts moved in rather low Earth orbits, rendering the areas
located near the celestial equator  periodically unobservable. At
the same time, for the CMB data such a sensitivity to the
equatorial coordinate system is impossible to explain in terms of
such a simple model, as the microwave background data were
obtained onboard the WMAP satellite, which rotates about the L2
Lagrangian point. Note that in our previous paper~\cite{cor_ecl}
we also found some correlations to be ``aware'' of the equatorial
system.
We do not rule out the possible contribution from the
Earth's magnetic field, which has a large extent and shows up as
large-scale correlations of the microwave background. However, we
do not understand the mechanism of such correlations.

Neither do we understand the possible interrelation between the
ecliptic plane and the positions of gamma-ray bursts. Here the
situation is reversed: the ecliptic features in the CMB data have
already been discussed recently~\cite{diego,dikarev,cor_ecl},
however, further studies are needed to understand what may link
gamma-ray evens and the ecliptic plane. One of the possible
hypotheses explains this feature as a selection effect due to the
fact that the observing instruments of the satellites are turned
away from the Sun, and this may put the plane of the ecliptic in a
special position.

Note also that the correlation properties of the CMB, which show
up in the ``near-Earth'' coordinate systems when the data with
``randomly'' distributed events are used, are indicative of the
non-Gaussian nature of the data for low-order multipoles, which
may be due either to the systematic effects, or to a hitherto
unexplored effect in the near-Earth space. We consider further
studies of this correlation to be of especially great interest
given that the new high-quality data are expected to be provided
by the Fermi and Planck missions.

\noindent
{\small
{\bf Acknowledgments}.
We are grateful to Valery Larionov (St. Petersburg State
University) for useful discussions of the results of this work. We
thank the NASA for making available the NASA Legacy Archive, from
where we adopted the WMAP data. We are also grateful to the
authors of the HEALPix\footnote{\tt
http://www.eso.org/science/healpix/} \cite{healpix} package, which
we used to transform the WMAP7 maps into the coefficients $a_{\ell
m}$. This work made use of the \mbox{GLESP\footnote{\tt
http://www.glesp.nbi.dk} \cite{glesp,glesp1}} package for the
further analysis of the CMB data on the sphere. This work was
supported by the Program for the Support of Leading Scientific
Schools of Russia (the School of S.~E.~Khaikin) and the Russian
Foundation for Basic Research (grant nos.~09-02-00298 and
\mbox{08-02-00486.} O.V.V. also acknowledges partial support from
the Foundation for the Support of Domestic Science (the program
``Young Doctors of Science of the
Russian Academy of Sciences'') and the Dynasty Foundation.  


\begin{thebibliography}{}

\bibitem{bepposax}
D.~Riccia, F.~Fioreb, and P.~Giommia,
Nuclear Physics B - Proc. Suppl.
{\bf 69}, 618 (1999).


\bibitem{batse}
W.~S.~Paciesas, C.~A.~Meegan, G.~N.~Pendleton, et al.
    \apjs~ {\bf 122}, 465 (1999),
astro-ph/9903205.


\bibitem{grb_apm}
L.~L.~R.~Williams  and N.~Frey,
\apj~ {\bf 583}, 594 (2003).

\bibitem{grb_vor}
A.~M\'esz\'aros, L.~G.~Bal\'azs, Z.~Bagoly, and P.~Veres,
arXiv:0906.4034

\bibitem{zse}
Ya.~B.~Zeldovich and R.~A.~Sunyaev.
Astrophys. Space Sci. {\bf 4}, 301 (1969).


\bibitem{swe}
R.~K.~Sachs and A.~M.~Wolfe,
\apj~ {\bf 147}, 73 (1967).

\bibitem{wmap7ytem}
N.~Jarosik, C.~L.~Bennett, J.~Dunkley, et al.,
{\apjs}, submitted (2010), arXiv:1001.4744.

\bibitem{wmapresults}
 C.~L.~Bennett, M.~Halpern,  G.~Hinshaw, ~et al.,
{\apjs} {\bf 148}, {1}	(2003), {astro-ph/0302207}.

\bibitem{cormap}
O.~V.~Verkhodanov, M.~L.~Khabibullina, and E.~K.~Majorova,
Astrophys. Bull.
 {\bf 64}, 263 (2009).

\bibitem{glesp2}
A.~G.~Doroshkevich, O.~B.~Verkhodanov, O.~P.~Naselsky, et al.,
arXiv0904.2517	(2009).

\bibitem{cor_ecl}
O.~V.~Verkhodanov and M.~L.~Khabibullina, Astrophys. Bull. {\bf 65}, accepted
(2010).

\bibitem{diego}
J.~M.~Diego, M.~Cruz, J.~Gonzalez-Nuevo,  et al.,
arXiv: 0901.4344 (2009).

\bibitem{dikarev}
V.~Dikarev, O.~Preuss, S.~Solanki, et al., 
 \apj~ {\bf 705}, 670 (2009).

\bibitem{healpix}  
K.~G\'orski, E.~Hivon, A.~J.~Banday, et al.,
{\apj} {\bf 622}, {759} (2005).

\bibitem{glesp} 
 A.~G.~Doroshkevich, P.~D.~Naselsky, O.~V.~Verkhodanov, et al.,
    {Int. J. Mod. Phys. D} {\bf 14}, {275} ({2003}),
{astro-ph/0305537}.

\bibitem{glesp1}
 O.~V.~Verkhodanov, A.~G.~Doroshkevich,	 P.~D.~Naselsky, et al.,
     \bsao~ {\bf  58}, 40 (2005).

\end{thebibliography}
\end{document}